\documentclass{lmcs} %%% last changed 2014-08-20
\pdfoutput=1

\usepackage{lastpage}
\lmcsdoi{22}{2}{7}
\lmcsheading{}{\pageref{LastPage}}{}{}%
{Mar.~22,~2025}{Apr.~17,~2026}{}


\usepackage{iftex}
\ifpdf
  \usepackage{underscore}         % Only needed if you use pdflatex.
  \usepackage[T1]{fontenc}        % Recommended with pdflatex
\else
  \usepackage{breakurl}           % Not needed if you use pdflatex only.
\fi

\usepackage{listings}
\usepackage{subfigure}
\usepackage{xspace}
\usepackage{graphicx}
\usepackage[utf8]{inputenc}
\usepackage{multirow}
\usepackage{enumitem}
\usepackage{rotating}
\usepackage{float}
\usepackage{longtable}

\usepackage{algpseudocode}
\usepackage{graphicx}
\usepackage{amsmath}
\usepackage{mathtools}
\usepackage{algorithm}
\usepackage{syntax}
\usepackage{amssymb}
\usepackage{array}
\usepackage[font=normalsize]{caption}
\usepackage{booktabs}
\newcommand{\bigB}[1]{\big[ #1 \big]}

\newcommand{\deq}{{\stackrel{\mathrm{def}}{=}}}
\newcommand{\defiff}{{\stackrel{\mathrm{def}}{\mIff}}}

\newcommand{\C}{{\mathcal{C}}}

\newcommand{\semiring}[5]{\big(#1, #2, #3, #4, #5\big)}
	\usepackage{stackrel}
\newcommand{\nAnd}{\;\mathrel{\wedge}\;}

\newcommand{\mImp}{\;\Longrightarrow\;}  
  
\newcommand{\mIff}{\;\Longleftrightarrow\;} 

\newcommand{\lnotation}[4]{
	\def\third:{#3} 
	\def\possiblyone:{} 
	\def\possiblytwo:{~}
	\def\possiblythree:{ }
	\def\divide{\;#1\hspace*{-0pt}( #2\; \mid: \; #4 \, )}
	\def\nodivide{\;#1\hspace*{-0pt}( #2\;\mid\; #3\;:\;#4 \, )}
	\ifx\third\possiblyone\divide
		\else\ifx\third\possiblytwo\divide
		\else \ifx\third\possiblythree\divide
		\else \nodivide\fi\fi\fi}

\newcommand{\biglnotation}[4]{
	\def\third:{#3} 
	\def\possiblyone:{} 
	\def\possiblytwo:{~}
	\def\possiblythree:{ }
	\def\divide{\;#1\hspace*{-0pt}\big( #2\; \mid: \; #4 \, \big)}
	\def\nodivide{\;#1\hspace*{-0pt}\big( #2\;\mid\; #3\;:\;#4 \, \big)}
	\ifx\third\possiblyone\divide
		\else\ifx\third\possiblytwo\divide
		\else \ifx\third\possiblythree\divide
		\else \nodivide\fi\fi\fi}

\newcommand{\bigglnotation}[4]{
	\def\third:{#3} 
	\def\possiblyone:{} 
	\def\possiblytwo:{~}
	\def\possiblythree:{ }
	\def\divide{\;#1\hspace*{-0pt}\bigg( #2\; \mid: \; #4 \, \bigg)}
	\def\nodivide{\;#1\hspace*{-0pt}\bigg( #2\;\mid\; #3\;:\;#4 \, \bigg)}
	\ifx\third\possiblyone\divide
		\else\ifx\third\possiblytwo\divide
		\else \ifx\third\possiblythree\divide
		\else \nodivide\fi\fi\fi}

\newcommand{\grieslnotation}[4]{
	\def\third:{#3} 
	\def\possiblyone:{} 
	\def\possiblytwo:{~}
	\def\possiblythree:{ }
	\def\divide{(#1 #2\; \mid : \; #4 \, )}
	\def\nodivide{(#1 #2\;\mid\; #3\;:\;#4 \, )}
	\ifx\third\possiblyone\divide
		\else\ifx\third\possiblytwo\divide
		\else \ifx\third\possiblythree\divide
		\else \nodivide\fi\fi\fi}
	\usepackage{color}
	\usepackage{etoolbox}
	\usepackage{soul}
\definecolor{darkred}{rgb}{0.75,0.0,0.0}
\definecolor{darkgreen}{rgb}{0.0,0.6,0.0}
\definecolor{darkblue}{rgb}{0.0,0.0,0.6}
\definecolor{darkcyan}{rgb}{0.0,0.6,0.6}
\definecolor{darkmagenta}{rgb}{0.6,0.0,0.6}
\definecolor{darkamber}{rgb}{1.0,0.5,0.0}
\definecolor{darkyellow}{rgb}{0.6,0.6,0.0}

\definecolor{lightred}{rgb}{1.0,0.9,0.9}
\definecolor{lightgreen}{rgb}{0.9,1.0,0.9}
\definecolor{lightblue}{rgb}{0.9,0.9,1.0}
\definecolor{lightcyan}{rgb}{0.8,1.0,1.0}
\definecolor{lightmagenta}{rgb}{1.0,0.8,1.0}
\definecolor{lightamber}{rgb}{1.0,0.8,0.0}
\definecolor{lightyellow}{rgb}{1.0,1.0,0.8}

\definecolor{webgreen}{rgb}{0,0.5,0}
\definecolor{webbrown}{rgb}{0.6,0,0}

\definecolor{grey}{rgb}{0.65,0.65,0.65}
\definecolor{purple}{rgb}{0.4,0,0.75}

\definecolor{burgundy}{rgb}{0.5, 0.0, 0.13}         % For states
\definecolor{darkcyan}{rgb}{0.0,0.6,0.6}            % For sync messages
\definecolor{darkpastelgreen}{rgb}{0.01, 0.75, 0.24}% For guarded transitions

\newcommand{\mynote}[2]{
	\ifstrequal{#1}{0}{\textcolor{darkamber}{#2}}{}%
  	\ifstrequal{#1}{1}{\textcolor{darkmagenta}{#2}}{}%
  	\ifstrequal{#1}{2}{\textcolor{darkcyan}{#2}}{}%
  	\ifstrequal{#1}{3}{\textcolor{darkgreen}{#2}}{}%
  	\ifstrequal{#1}{5}{\textcolor{darkblue}{#2}}{}%
  	\ifstrequal{#1}{8}{\textcolor{burgundy}{#2}}{}
}
\newcommand{\todiscuss}[2]{
	\ifstrequal{#1}{0}{\textcolor{darkamber}{\textit{\textbf{TO DISCUSS}: #2}}}{}%
  	\ifstrequal{#1}{1}{\textcolor{darkmagenta}{\textit{\textbf{TO DISCUSS}: #2}}}{}%
  	\ifstrequal{#1}{2}{\textcolor{darkcyan}{\textit{\textbf{TO DISCUSS}: #2}}}{}%
  	\ifstrequal{#1}{3}{\textcolor{darkgreen}{\textit{\textbf{TO DISCUSS}: #2}}}{}
}
\newcommand{\todo}[2]{
	\ifstrequal{#1}{0}{\textcolor{darkamber}{\textbf{\underline{TO DO}}: #2}}{}%
  	\ifstrequal{#1}{1}{\textcolor{darkmagenta}{\textbf{\underline{TO DO}}: #2}}{}%
  	\ifstrequal{#1}{2}{\textcolor{darkcyan}{\textbf{\underline{TO DO}}: #2}}{}%
  	\ifstrequal{#1}{3}{\textcolor{darkgreen}{\textbf{\underline{TO DO}}: #2}}{}
}
\newcommand{\toaddress}[2]{
	\ifstrequal{#1}{0}{\noindent\textcolor{darkamber}{$\bigstar$~\textbf{#2}}\\}{}%
  	\ifstrequal{#1}{1}{\noindent\textcolor{darkmagenta}{$\bigstar$~\textbf{#2}}\\}{}%
  	\ifstrequal{#1}{2}{\noindent\textcolor{darkcyan}{$\bigstar$~\textbf{#2}}\\}{}%
  	\ifstrequal{#1}{3}{\noindent\textcolor{darkgreen}{$\bigstar$~\textbf{#2}}\\}{}
}
\newcommand{\eg}{\textrm{e.g.,}\@\xspace}
\newcommand{\ie}{\textrm{i.e.,}\@\xspace}

\newcommand{\etal}{\textrm{et~al.}\@\xspace}

\newcommand{\opt}[1]{\mathit{opt}\bigB{#1}}
\newcommand{\refines}{\sqsubseteq}
\newcommand{\requires}[3]{#1 \xrightarrow[]{#3} #2}

\newcommand{\ravenclaw}{\textsl{Ravenclaw}\@\xspace}

\newtheorem{definition}[thm]{Definition} %FIXED to LMCS general counter

\newlist{tabitemize}{itemize}{1}
\setlist[tabitemize]{label=\textbullet, 
                     leftmargin=*,
                     nosep, 
                     before=\begin{minipage}[t]{\hsize}\raggedright, 
                     after=\end{minipage}}
\begin{document}

% If the title is longer than 55 characters, then specify a shorter running title as the optional argument to \title. The running title should be roughyl at most 55 characters:
\title[A Scalable Game-Theoretic Approach for Selecting Security Controls]{A Scalable Game-Theoretic Approach for Selecting Security Controls from Standardized Catalogues\rsuper*}

\thanks{\lsuper*A preliminary version of this article was presented at the 15th International Symposium on Games, Automata, Logics, and Formal Verification (GandALF 2024)~\cite{GandALF2024}. In this version, we extend the article by applying the proposed approach to a much larger use case system with the use of a custom tool.
}	%optional

% affiliations are numbered automatically with a, b, c (see below)
% use the optional argument to indicate the affiliation(s) of each author
% omit the argument if there is only one author, or only one affiliation
\author[D. L\'{e}veill\'{e}]{Dylan L\'{e}veill\'{e}\lmcsorcid{0009-0001-3606-5370}}[]
\author[J. Jaskolka]{Jason Jaskolka\lmcsorcid{0000-0001-6316-3040}}[]

% affiliation 1 (automatically numbered a)
\address{Carleton University, Canada}	%optional
% write emails for all authors having that affiliation
\email{dylan.leveille@carleton.ca, jason.jaskolka@carleton.ca}  %optional

% affiliation 2 (automatically numbered b)
% \address{University 2, address2}	%optional
% \email{name2@email2}  %optional

%% etc.

%% required for running head on odd and even pages, use suitable
%% abbreviations in case of long titles and many authors:

%%%%%%%%%%%%%%%%%%%%%%%%%%%%%%%%%%%%%%%%%%%%%%%%%%%%%%%%%%%%%%%%%%%%%%%%%%%

%% the abstract has to PRECEDE the command \maketitle:
%% be sure not to issue the \maketitle command twice!

\begin{abstract}
    
Selecting the combination of security controls that will most effectively protect a system's assets is a difficult task. If the wrong controls are selected, the system may be left vulnerable to cyber-attacks that can impact the confidentiality, integrity, and availability of critical data and services. In practical settings, as standardized control catalogues can be quite large, it is not possible to select and implement every control possible. Instead, considerations, such as budget, effectiveness, and dependencies among various controls, must be considered to choose a combination of security controls that best achieve a set of system security objectives. In this paper, we present a game-theoretic approach for selecting effective combinations of security controls based on expected attacker profiles and a set budget. The control selection problem is set up as a two-person zero-sum one-shot game. Valid control combinations for selection are generated using an algebraic formalism to account for dependencies among selected controls. Using a software tool, we apply the approach on a fictional Canadian military system with Canada's standardized control catalogue, ITSG-33. Through this case study, we demonstrate the approach's scalability to assist in selecting an effective set of security controls for large systems. The results illustrate how a security analyst can use the proposed approach and supporting tool to guide and support decision-making in the control selection activity when developing secure systems of all sizes.
\end{abstract}

\maketitle

% Paper Body
\section{Introduction}
\label{sec:introduction}
% Begin Section

With computers becoming more interconnected than ever, there emerges an even greater need to secure computer systems and to effectively manage security risks. Security risks are mitigated by the implementation of a set of security controls. A \textit{security control} refers to a safeguard or countermeasure prescribed for an information system or an organization designed to protect the confidentiality, integrity, and availability of its information and to meet a set of defined security requirements~\cite{NIST-800-160v2r1}.  
\noindent %FIXED margins
\textit{Control selection} is an activity commonly found as part of a risk management process~\cite{ISO-31000}, a systems engineering process~\cite{NIST-800-160v2r1}, the Risk Management Framework~\cite{NIST-RMF}, the Cybersecurity Framework~\cite{NIST-CSF}, or the Privacy Framework~\cite{NIST-PF}.
Control selection involves selecting and documenting the security controls necessary to protect the information system and organization commensurate with risk to organizational and system operations and assets, individuals, other organizations, and the nation~\cite{NIST-RMF}. 

During the control selection activity, security analysts typically select security controls from standardized security control catalogues, such as NIST~SP~800-53~\cite{NIST-800-53r5}, ITSG-33~\cite{ITSG-33}, ISO~27002~\cite{ISO-27002}, CIS Critical Security Controls~\cite{cisControls}, and MITRE D3FEND\textsuperscript{\texttrademark}~\cite{DEFEND}, among others. However, selecting combinations of controls from these catalogues can be difficult for several reasons. 
First, these control catalogues are large, and many possible controls could be selected to mitigate the risks identified for a given system. In practical settings, it is not possible to select and implement every control possible. Considerations such as budget, effectiveness, and dependencies among various controls, must be considered to choose a combination of security controls that best achieve a set of system security objectives. 
Second, control selection is largely a human-oriented activity. The dynamics between security analysts (defenders) strategizing to protect critical systems and assets and achieve a set of security objectives, and attackers aiming to impact critical systems and assets and violate those same security objectives must be considered when deciding on the most effective and cost-efficient combination of security controls. 
Although numerous optimization-based solutions are adept at accounting for various properties of the controls themselves, they fail to capture the human element that is inherently part of the control selection activity. 

To address the above-mentioned challenges, we propose a game-theoretic approach for security control selection. The human aspects of the control selection problem, as well as the large space of possible control combinations, and their dependencies and constraints, lend themselves well to an application of game theory. Specifically, we set up a two-person zero-sum one-shot game which is played by a security analyst. The analyst selects their strategy based on an attacker profile, characterized by the expected targeted assets and security objectives. Each analyst strategy corresponds to a combination of security controls from a chosen control catalogue that can achieve the security objectives. Valid control combinations are generated using an algebraic formalism (akin to product family algebra~\cite{Hofner2011}) to account for dependencies among selected controls. The outcome of the game is a combination of suggested security controls that can effectively defend against the considered attacker profile. To demonstrate the approach's scalability towards large systems, we apply the approach on an illustrative Canadian military system using a previously proposed tool~\cite{FPS2024}.  

The rest of this paper is organized as follows. Section~\ref{sec:relatedwork} provides an overview of existing works on the topic of control selection and of game theory applications in cybersecurity. Section~\ref{sec:approach} presents the proposed game-theoretic approach for control selection. Section~\ref{sec:csat} presents a previously proposed tool that automates the proposed approach. Section~\ref{sec:example} provides a large illustrative example in which the proposed approach is applied using the tool from Section~\ref{sec:csat}. Section~\ref{sec:discussion} discusses the benefits and potential limitations of the proposed approach. Lastly, Section~\ref{sec:conclusion} concludes and briefly discusses future work.

% End Section

\section{Related Work}
\label{sec:relatedwork}
% Begin Section
In this section, we present existing works that propose solutions for assisting with control selection. Additionally, we discuss the lack of scalability of these approaches, thus highlighting their impracticality for large real-world systems.

\subsection{Control Selection Approaches}
Many existing approaches to support the security control selection activity are based on setting and solving optimization problems. For example, for each considered control, Yevseyeva \etal~\cite{Yevseyeva2015} assign a probability of ``survival'' for each possible threat (\ie the probability that the threat persists in the presence of the control). Probabilities are also assigned for the expected loss of successful attacks. The goal of the proposed approach is to minimize this expected loss, under constraints such as cost and system resources. Similarly, Almeida and Respício~\cite{Almeida2018} also assign probabilities to controls based on their expected performance in mitigating certain vulnerabilities. For the proposed approach, the goal is to find the optimal controls for the system that will minimize an objective function accounting for both loss and cost. Many such optimization approaches have been proposed, differing only by the metrics used for optimization and their specific optimization techniques. This includes the works by Tsiodra~\etal~\cite{Tsiodra2021}, Schmidt~\etal~\cite{Schmidt2021}, and Sarala~\etal~\cite{Sarala2016}. A different approach was proposed by Dewri \etal~\cite{Dewri2007} where systems are modelled as trees, in which the leaf nodes represent possible attacks. Controls therefore mitigate one or many leaf nodes. With the attack impact, attack frequency, and cost of each control known, the optimal controls can be found by optimization. A similar tree-like approach was also proposed by Park and Huh~\cite{Park2020}. Using an Attack-Defense tree, in which attacker motivations and defender mitigations are modeled as a tree, is also a popular approach for assisting with control selection~\cite{kordy2014attack,fila2020exploiting,ji2016attack}. Although these trees can be modeled in various ways, optimal solutions are obtained through optimization.  Lastly, Shahpasand~\etal~\cite{Shahpasand2015} and Uuganbayar~\etal~\cite{Uuganbayar2021} have both developed custom algorithms that solve control selection as an optimization problem.  While optimization-based approaches can account for important considerations and constraints such as cost and effectiveness, they depend heavily on assumptions about probabilities for threat likelihoods or control success rates. Such probabilities are not likely to be accurately known in a practical setting. 

Several other approaches for control selection that are not based on optimization have also been proposed.
Bettaieb \etal~\cite{Bettaieb2020} presented an approach where a machine learning model is trained with historical data from previous security assessments to make predictions using certain features of interest from a given security assessment to determine optimal controls. However, using historic data to determine how to protect a system has several limitations as every system is unique and may operate in widely different environments. In another work, Kiesling \etal~\cite{Kiesling2016} proposed a simulation-based approach to determine the optimal controls for a system. To do this, expected attacks are simulated on different components of the system using different possible control combinations to find the optimal ones. This approach is noteworthy as it simply uses the properties of the controls and of the current system (such as different threats) to find the most optimal control combinations and does not depend on any probabilities.

Although the approaches presented above may be suitable for solving the control selection problem, they fail to account for the human behavioural aspect of an attacker, opting instead to categorize potential attackers as a series of possible attacks. Additionally, the main issue with many of these approaches is the need to define probabilities for threat likelihoods or control success rates. Such probabilities are not likely to be known in any realistic scenario. These works also fail to acknowledge possible dependencies between the controls. In contrast to existing work, the proposed approach aims to leverage game theory to address the shortcomings of current control selection approaches by placing a central focus on possible attacker behaviours, while also considering the dependencies and constraints that limit the selection of certain combinations of security controls to effectively mitigate the threats to a system. 

\subsection{Scalability of Control Selection}
While existing approaches fail to capture the human nature of this problem, they lend themselves well to being automated as many of them simply involve solving an optimization problem. In fact, most of these works mention the necessity of automation to apply their approaches practically. For example, Almeida and Respício~\cite{Almeida2018} developed an Excel tool to automate their approach. However, solutions are generated through brute-force optimization. Brute-force methods are also used to automate the approach by Park and Huh~\cite{Park2020}. While brute-force optimization is functional, given the numerous constraints that exist in control selection, it is not scalable for cases in which a large number of controls are considered. 
% This is because the control selection problem is a variant of the knapsack problem.
Although the works of Shahpasand~\etal~\cite{Shahpasand2015} avoid brute-force optimization through the development of a custom algorithm, they provide no details on the computational complexity of their algorithm. Similarly, while Uuganbayar~\etal~\cite{Uuganbayar2021} provides experimental showing that their algorithm finds results within five to ten minutes, the computational complexity of the algorithm is not provided.

% Several works have explored the use of game theory for addressing cybersecurity challenges. 
% For example, Nassar \etal~\cite{Nassar2021} proposed a technique which focused on evaluating a system's network security with the help of a game model. Smith \etal~\cite{Smith2017} used game theory to verify the security of hardware designs. Wang \etal~\cite{Wang2010} presented a network attack-defence game to help secure a computer network. However, game theory has yet to be utilized for security control selection.

The literature shows that there is a lack of research in the automation of control selection approaches as the computational complexity of this problem is overlooked. 
% Proposed approaches generate solutions inefficiently when a large subset of control options is considered, rendering them impractical for use in large-scale modern computer systems. 
% Additionally, as the approaches rely purely on optimization and neglect the human elements of this problem, their solutions may not be useful~\cite{GandALF2024}. 
Additionally, as existing approaches mostly rely on optimization, an automated game-theoretic approach for control selection has yet to be developed. Although the approach automated in this work is based on game theory, the constraints inherent in control selection necessitates that a custom algorithm be developed for efficient automation. Lastly, this work is the first to automate a control selection approach that considers control dependencies.
% End Section

\section{The Proposed Approach}
\label{sec:approach}
\begin{figure}[ht!]
    \centering
    \includegraphics[width=\textwidth]{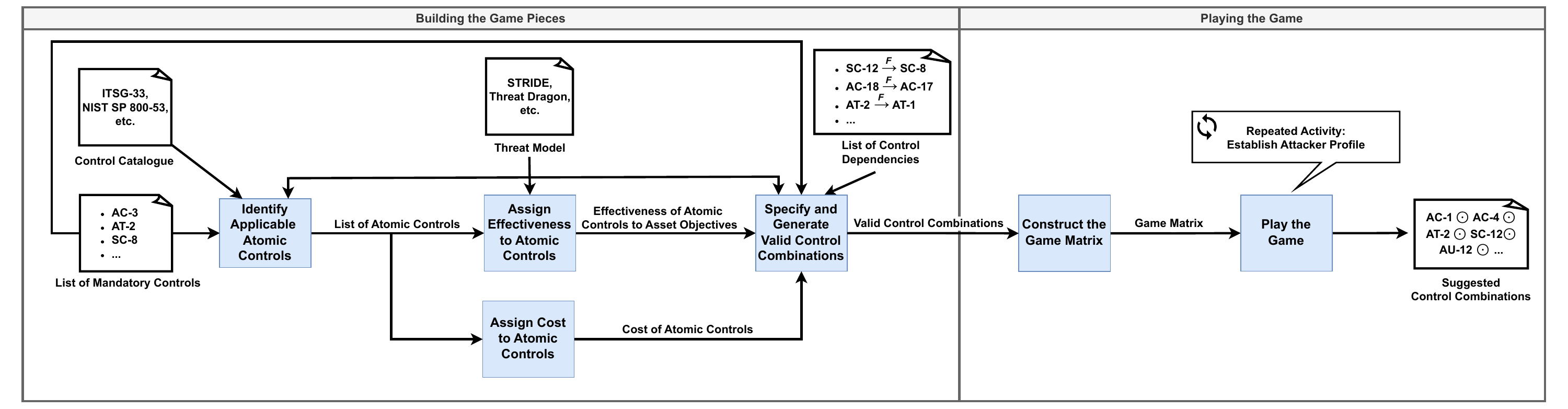}
    \caption{An overview of the game-theoretic approach for security control selection}
    \label{fig:propMeth}
\end{figure}

% Begin Section
In this section, we present our game-theoretic approach for security control selection. An overview of the approach is shown in Figure~\ref{fig:propMeth}.
The approach consists of two main stages shown as swim lanes and six steps shown in blue. 
All steps are to be conducted by a security analyst. A detailed description of each step of the proposed approach is provided in the sections below. 

\subsection{Identify Applicable Atomic Controls}

Before the approach can be applied, the analyst must be provided with a security control catalogue, such as NIST SP 800-53~\cite{NIST-800-53B} or ITSG-33~\cite{ITSG-33}, from which security controls for the system will be considered.  The approach begins with the security analyst identifying applicable atomic controls from this given control catalogue. In our context, the \textit{atomic controls} are the smallest (indivisible) security controls that can be selected from a control catalogue. We say that a control is \textit{applicable} to a system if it could provide any form of protection from the threats to the assets of the system. 

We assume that a list of threats and assets are available to the security analyst in the form of a threat model. A threat model is defined as ``a structured representation of all the information that affects the security of an application''~\cite{Drake}. Threat models typically include identified system threats and their impact on the assets within the system~\cite{Drake,NIST-800-154}. 
The threat model can be obtained by applying a well-known threat modelling methodology such as STRIDE~\cite{STRIDE} or PASTA~\cite{PASTA}.

To determine the applicability of an atomic control, the security analyst must carefully consider each atomic control from the given control catalogue and decide if the control can mitigate the identified threats to the assets. Additionally, certain organizational needs or standards and regulations for the system's application domain may require that specific security controls be present in the system. The security analyst must therefore ensure that these \textit{mandatory controls} are included as part of the set of applicable controls identified. 
This is a manual process. However, it should be noted that the effort required for this activity is reasonable as security control catalogues are typically separated by control families which help guide an analyst in finding suitable controls~\cite{Rouland2023ab}. Instructions and guidance for performing this task is well documented by ISO~27005~\cite{ISO-27005} and NIST~SP~800-53B~\cite{NIST-800-53B}.

\textit{At the end of this step, the analyst will have a set of applicable atomic controls for the system.} Note that the combination of suggested controls found by applying the proposed approach will be a subset of the controls gathered in this initial step.

\subsection{Assign Effectiveness to Atomic Controls}
\label{sec:approach:eff}

For each identified atomic control, the analyst proceeds by assigning an effectiveness of the control at satisfying each \textit{security objective} on each asset in the system. Security objectives represent the security needs of the assets on the system, such as confidentiality, integrity, and availability~\cite{NIST-1800-26}. These objectives are normally included as part of the threat impacts described in the threat model. It is important to remember that the goal of the proposed approach is to create a game. In every game, there needs to be strategies, and payoffs defined for each strategy. Assigning the effectiveness of each atomic control therefore defines the payoffs of each atomic control in the game. 

To perform this step of the approach, the atomic payoff matrix presented in Table~\ref{tab:atomicPayoff} must be completed. The rows represent each atomic control that was identified in the previous step (denoted $C_{1},\dots,C_{N}$). The columns represent the security objectives for each asset (denoted $O_{1},\dots,O_{M}$). We expect the analyst to assign a value between 0 and 1 in each cell of this matrix. A value of 0 means that the atomic control is not effective at satisfying the specified objective for an asset, while a value of 1 means that the atomic control is completely effective at satisfying the specified objective for an asset. Each payoff value is therefore normalized. Provided that the rating scheme is selected and used consistently throughout the approach, the analyst is free to choose any method for assigning the effectiveness values for the atomic payoff matrix. For example, the analyst may choose to use a quantitative approach as in the Defect Detection and Prevention (DDP) risk reduction strategy developed by NASA~\cite{Feather2005}, or they may alternatively choose to use a qualitative rating mapped to quantitative values as in~\cite{Liu2011aa}.

\begin{table}[ht!]
\caption{General form of the atomic payoff matrix}
\label{tab:atomicPayoff}
    \centering
    \small
    \begin{tabular}{|c|c|c|c|c|c|c|c|c|c|c|c|c|c|c|}
    \cline{2-11}
    \multicolumn{1}{c|}{} & \multicolumn{3}{c|}{\textit{Asset 1}} & \multicolumn{3}{c|}{\textit{Asset 2}} & \multicolumn{1}{c|}{\dots} & \multicolumn{3}{c|}{\textit{Asset X}}\\
    %  \cline{6-8}
    %   \multicolumn{5}{c|}{} & \multicolumn{5}{c|}{numbers2} \\
    \cline{2-11}
    \multicolumn{1}{c|}{} & $O_{1}$ & \dots & $O_{M}$ & $O_{1}$ & \dots & $O_{M}$ & \dots & $O_{1}$ & \dots & $O_{M}$ \\
    \hline
    $C_{1}$ &  &  &  & & & & &  & &\\
    \cline{1-11}
    % $C_{2}$ &  &  &  & & & & &  & &\\
    % \cline{1-11}
    \vdots &  &  &  & & & & &  & &\\
    \cline{1-11}
    $C_{N}$ &  &  &  & & & & &  & &\\
    \hline
    \end{tabular}
\end{table}

\textit{At the end of this step, the analyst will have the effectiveness of each applicable atomic control for satisfying each security objective on each asset in the system. }

\subsection{Assign Cost to Atomic Controls}
\label{sec:approach:cost}

In practical settings, cost or time constraints limit how many controls can be part of a system; if there are too many controls they may exceed a certain budget or cannot be implemented in reasonable time. In fact, without such constraints, there could technically be no limitations on the number of controls that can be selected for a system, and the best solution would be to select them all. 

At the same time as assigning effectiveness, the analyst will also need to assign a cost for each identified atomic control. We expect the analyst to assign a cost from the set of real numbers $\mathbb{R}$. The units for cost could be represented as dollars, thousands of dollars, or any other form of currency as long as the same units are consistently used for all cost values. Furthermore, no units could be used if desired. Without units, costs simply represent an implementation effort.

\textit{After this step, the analyst will have the cost associated with each applicable atomic control. }

\subsection{Specify and Generate Valid Control Combinations}

Given a set of applicable atomic controls, the analyst needs to specify and generate the set of valid control combinations that satisfies their constraints. To formally capture these constraints, we have decided to use an algebraic specification based on product family algebra~\cite{Hofner2011} to specify and generate valid combinations of security controls.

Product family algebra extends the mathematical notions of semirings to describe and manipulate product families. A semiring is an algebraic structure~$\semiring{S}{+}{\cdot}{0}{1}$ consisting of a set $S$ with a commutative and associative binary operator $+$ and an associative binary operator $\cdot$. An element $0 \in S$ is the identity element with respect to $+$, while an element $1 \in S$ is the identity element with respect to $\cdot$. Additionally, $\cdot$ distributes over $+$ and element $0$ annihilates $S$ with respect to $\cdot$. A semiring is commutative if $\cdot$ is commutative and a semiring is idempotent if $+$ is idempotent.

 For ease of presentation, we recast the vocabulary of product family engineering into the vocabulary of security controls by first defining a security control algebra to express families of security control combinations generated from a set of atomic controls. 

\begin{definition}[Security Control Algebra]
\label{def:controlAlgebra}
    A \emph{security control algebra} is a commutative idempotent semiring~$\C\ \deq \semiring{C}{\oplus}{\odot}{0}{1}$ where each element of the semiring $c \in C$ is a security control family.
\end{definition}
In a security control algebra, the operator $\oplus$ is interpreted as a choice between two security control families and the operator $\odot$ is interpreted as a mandatory composition of two security control families\footnote{When the context is clear, we omit the mandatory composition operator $\odot$ when specifying security control algebra terms.}. The element $0$ represents a non-implementable security control combination that cannot exist and the element $1$ represents the empty security control combination which has no controls.
A security control family is called a \textit{security control combination} if it is indivisible with regard to the choice operator $\oplus$. Additionally, it is called a \textit{proper security control combination} if  $c \neq 0$. A security control combination is an \textit{atomic control} if is it is indivisible with regard to the mandatory composition operator~$\odot$.
Optional controls are expressed as a choice between the controls and the empty security control combination $1$. A list of optional controls $c_1,\dots,c_n$ is denoted by $\opt{c_1,\dots,c_n} \deq (c_1 \oplus 1) \odot \dots \odot (c_n \oplus 1)$. 

For two security control families $c_1$ and $c_2$ in a security control algebra, the \textit{refinement relation} ($\refines$) is defined as $c_1 \refines c_2 \defiff \lnotation{\exists}{c_3}{}{c_1 \leq c_2 \odot c_3}$ where $\leq$ is the natural semiring order (\ie $c_1 \leq c_2 \defiff c_1 \oplus c_2 = c_2$). 
To specify constraints, such as dependencies between controls, we use the requirement relation. 

\begin{definition}[Requirement Relation~\cite{Hofner2011}]
\label{def:reqRelation}
    For elements $c_1, c_2, c_3, c_4$ and security control combination $x$ in a security control algebra, the \textit{requirement relation} ($\rightarrow$) is defined inductively as:
    \begin{eqnarray*}
        \requires{c_1}{c_2}{x} &\defiff& x \refines c_1 \mImp x \refines c_2 \\
        \requires{c_1}{c_2}{c_3 \oplus c_4} &\defiff& \requires{c_1}{c_2}{c_3} \nAnd \requires{c_1}{c_2}{c_4}
    \end{eqnarray*}
\end{definition}
For elements $c_1, c_2$ and $x$, the requirement relation $\requires{c_1}{c_2}{x}$ can be read as ``$c_1$ requires $c_2$ within $x$.''

With this setting, all security control combinations can be specified algebraically by expressing the mandatory and optional controls as terms of a security control algebra along with requirement relations describing control dependencies.

The resulting specification serves as the basis for generating all possible proper security control combinations. However, not all control combinations are possible as some may exceed our defined budget. To make this determination we first define how to calculate the cost of a proper security control combination. In what follows, let $P \subseteq C$ be the set of all proper security control combinations in a security control algebra $\C$.

\begin{definition}[Cost of a Proper Security Control Combination]
\label{def:costInduction}
  The cost of a proper security control combination $\textit{Cost}: P \rightarrow \mathbb{R}$ is a function defined inductively for any proper security control combinations $a, b \in P$ in a security control algebra $\C$ as:
    \begin{eqnarray*}
        \mathit{Cost}(1) &=& 0\\
        \mathit{Cost}(a) &=& G(a)\ \text{if $a$ is atomic} \\
        \mathit{Cost}(a \odot b) &=& \mathit{Cost}(a) + \mathit{Cost}(b)
    \end{eqnarray*}
    where $G$ is a function that returns the cost assigned to an atomic control (see Section~\ref{sec:approach:cost}).    
\end{definition}

Now that we can compute the cost of a proper security control combination, we determine the set of valid security control combinations. A \textit{valid security control combination} is a proper security control combination that does not exceed the prescribed cost budget. The validity of a control combination is formalized in the following rule.
\begin{definition}[Budget Rule]
\label{def:rule}
    For any $p \in P$ and budget $B$:
    \begin{equation*}
        \mathit{Valid}(p) \mIff \mathit{Cost}(p) \leq B
    \end{equation*}
\end{definition}

\textit{After this step, the analyst will have a set of valid security control combinations that satisfy the prescribed budget.} These valid security control combinations become the strategies that an analyst can select when playing the game.

\subsection{Construct the Game Matrix}

In this step, the analyst constructs the game matrix. The general form of the game matrix can be seen in Table~\ref{tab:basicGameMatrix}. The rows represent the valid security control combinations found from the last step (denoted $\mathit{Combo}_{1},\dots,\mathit{Combo}_{N}$). The columns represent the security objectives for each asset (denoted $O_{1},\dots,O_{M}$). Note that the game matrix is identical in style to that of the atomic payoff matrix (see Table~\ref{tab:atomicPayoff}). The game matrix simply has control combinations as rows rather than atomic controls. 
In the game, the strategies of the security analyst will be the valid security control combinations, while the strategies of the attacker will be each security objective that could be violated on every asset. 

\begin{table}[ht!]
\caption{General form of the game matrix}
\label{tab:basicGameMatrix}
    \centering
    \small
    \begin{tabular}{l|lll|lll|l|lll|}
    \cline{2-11}
                             & \multicolumn{3}{c|}{\textit{Asset 1}}                            & \multicolumn{3}{c|}{\textit{Asset 2}}                            & \multicolumn{1}{c|}{\dots} & \multicolumn{3}{c|}{\textit{Asset X}}                            \\ \cline{2-11} 
                             & \multicolumn{1}{l|}{$O_{1}$} & \multicolumn{1}{l|}{\dots} & $O_{M}$ & \multicolumn{1}{l|}{$O_{1}$} & \multicolumn{1}{l|}{\dots} & $O_{M}$ & \dots                     & \multicolumn{1}{l|}{$O_{1}$} & \multicolumn{1}{l|}{\dots} & $O_{M}$ \\ \hline
    \multicolumn{1}{|l|}{$\mathit{Combo}_1$} & \multicolumn{1}{l|}{}   & \multicolumn{1}{l|}{}    &    & \multicolumn{1}{l|}{}   & \multicolumn{1}{l|}{}    &    &                          & \multicolumn{1}{l|}{}   & \multicolumn{1}{l|}{}    &    \\ \hline
    % \multicolumn{1}{|l|}{$\mathit{Combo}_2$} & \multicolumn{1}{l|}{}   & \multicolumn{1}{l|}{}    &    & \multicolumn{1}{l|}{}   & \multicolumn{1}{l|}{}    &    &                          & \multicolumn{1}{l|}{}   & \multicolumn{1}{l|}{}    &    \\ \hline
    \multicolumn{1}{|c|}{\vdots}   & \multicolumn{1}{l|}{}   & \multicolumn{1}{l|}{}    &    & \multicolumn{1}{l|}{}   & \multicolumn{1}{l|}{}    &    &                          & \multicolumn{1}{l|}{}   & \multicolumn{1}{l|}{}    &    \\ \hline
    \multicolumn{1}{|l|}{$\mathit{Combo}_N$} & \multicolumn{1}{l|}{}   & \multicolumn{1}{l|}{}    &    & \multicolumn{1}{l|}{}   & \multicolumn{1}{l|}{}    &    &                          & \multicolumn{1}{l|}{}   & \multicolumn{1}{l|}{}    &    \\ \hline
    \end{tabular}
\end{table}

Each outcome in a game is tied to a payoff~\cite{Straffin1993}. In our game, the payoffs are represented from the perspective of the analyst and represent the effectiveness of the security control combinations towards every asset's security objectives. Just as cost was defined inductively, we can define a proper control combination’s effectiveness towards an asset's security objective in a similar manner.

\begin{definition}[Effectiveness of a Proper Security Control Combination]
\label{def:effInduction}

    The effectiveness of a proper security control combination towards an asset's security objective $\mathit{Eff}: P \rightarrow \mathbb{R}$ is a function defined inductively for any proper security control combinations $a, b \in P$ in a security control algebra $\C$ as:
    \begin{eqnarray*}
        \mathit{Eff}(1) &=& 0\\
        \mathit{Eff}(a) &=& E(a)\ \text{if $a$ is atomic} \\
        \mathit{Eff}(a \odot b) &=& 1 - (1 - \mathit{Eff}(a))( 1- \mathit{Eff}(b) )
    \end{eqnarray*}
    where $E$ is a function that returns the effectiveness assigned to an atomic control for an asset's security objective (see Section~\ref{sec:approach:eff}). 
\end{definition}

With Definition~\ref{def:effInduction}, the payoff values in the game matrix can be calculated. Note that the calculation of the effectiveness of a security control combination is inspired from the combined effectiveness calculation as part of NASA's DDP approach~\cite{Feather2005}, in which effectiveness is calculated by subtracting the compounded ineffectiveness of each control in the combination from complete effectiveness (i.e., 1).

\textit{After this step, the analyst will have the game matrix so that they can proceed to play the game.}  

\subsection{Play the Game}

The game is a \textit{two-person zero-sum one-shot game}. The game is played by \textit{two persons}: the security analyst and the attacker. The attacker may embody one or multiple entities, but acts as a unified adversary. The goal of the security analyst is to select the security control combination that will best protect the security objectives for the assets they believe will be targeted by the attacker. Only one security control combination can be selected, hence it is a \textit{one-shot} game. On the other hand, the goal of the attacker is to attack assets and violate corresponding security objectives. 
An attacker could attack one or many assets and violate one or more objectives from a series of attacks. Regardless, an attacker will select which assets and objectives they will target and will commit to attacking the selected assets and objectives.
The attacker will naturally prefer attacking assets which are not properly defended, \ie those for which there are minimally effective security controls. The effectiveness values in the game matrix (payoffs) do not directly correlate to a loss to the attacker. However, it is easy to see that the higher the values, the more difficult it is for an attacker to conduct a successful attack leading to corresponding security objective violations. Therefore, what the security analyst gains in effectiveness is what the attacker loses in their ability to successfully conduct their attack; hence, it is a \textit{zero-sum} game. Note that this game is strictly non-cooperative; the analyst and attacker are competing directly and would never want to cooperate.

Using the game matrix, the analyst must select a strategy (\ie a valid security control combination) to play that  will best protect the system assets and security objectives that they believe are most important. To do this, an analyst must establish the expected attacker profile. An \textit{attacker profile} is an expected set of the assets and corresponding security objectives targeted by the attacker. One can imagine different classes of attackers having different capabilities, and different targets, thereby establishing different attacker profiles. In the context of a game, an attacker profile corresponds to guessing the attacker strategy so that it can be defended. This consideration of the dynamics of the analyst and the attacker strategies in this game is what differentiates it from existing security control selection approaches.  

It is impossible to know exactly which security objectives on which assets will be attacked, so assumptions must be made. One way to do this is to determine where most of the critical information flows in the system and which assets may be prone to more attacks (\ie have more expected threats). The combination of these ideas can help localize assets that are more attractive for attacks, and therefore puts the security objectives of these assets at higher risk of violation. Another way to do this is to consider the risk to each asset and corresponding security objectives for the identified threats to the system (which we consider known to the analyst). In this case, prioritizing defence of assets and security objectives targeted by  high risk threats may be a good approach. Regardless, once the attacker profile is determined, then the suggested strategy (\ie the most effective security control combination) can be found.

\begin{figure}[ht!]
    \centering
    \centerline{\includegraphics[width=\textwidth]{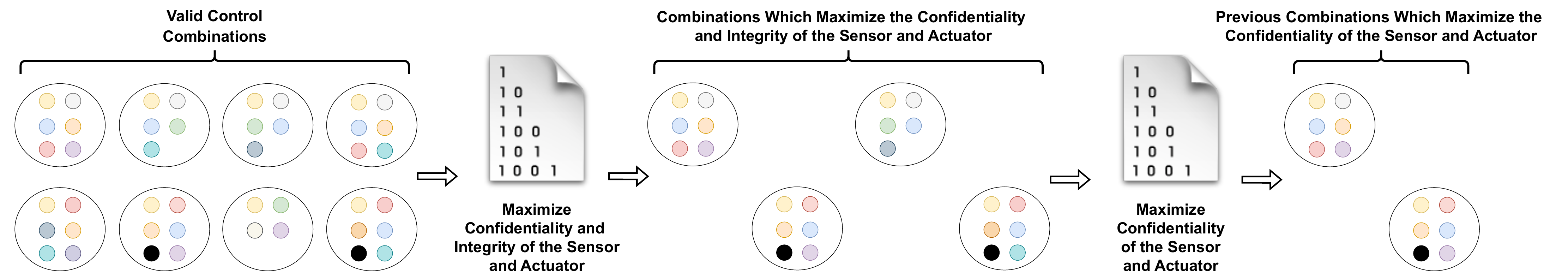}}
    \caption{Finding the suggested controls for an attacker profile with multiple ordered attacker objectives}
    \label{fig:combinedPersp}
\end{figure}

Regardless of the approach taken to establish the attacker profile, it will articulate the objectives that are expected to be violated by an attacker. For this work, we establish an attacker profile by considering and prioritizing different attacker objectives. Attacker objectives correspond to a set of security objectives for some assets that are equally expected to be targeted by an attacker. Within an attacker profile, several attacker objectives may be prioritized according to their perceived likelihood of being targeted by the attacker to obtain a priority order for the objectives. For example, the security analyst could establish an attacker profile in which the attacker has two ordered attacker objectives: (1) to target the confidentiality of two specific assets equally, and (2) to target the integrity of two other assets equally. The security analyst may consider as many attacker objectives as they desire when developing an attacker profile. The suggested analyst strategies for an attacker profile will be those which maximize the \textit{total effectiveness} across each attacker objectives (\ie the sum of the effectiveness returned by Definition~\ref{def:effInduction} for the security objectives in the attacker objectives is maximized in the priority order). To better understand this concept, an example of the strategies found by playing the game with an attacker profile with two ordered attacker objectives is visualized in Figure~\ref{fig:combinedPersp}.

In this example, there are initially eight valid security control combinations. Each security control combination has a unique set of controls (denoted by the different coloured dots in the figure). Only two  assets exist in this system; a Sensor and an Actuator. The attacker profile has two ordered attacker objectives: (1) the confidentiality and integrity of the Sensor and the Actuator and then (2) the confidentiality of the Sensor and the Actuator. From all valid control combinations, the control combinations which maximize the first set of attacker objectives is found, yielding four different combinations. From these four combinations, the combinations which maximize the second set of attacker objectives is found, yielding two control combinations. As there are no more ordered attacker objectives, the two resulting control combinations are considered equally valid, and represent the suggested strategies. Note that since the suggested strategies are derived through a series of maximization problems, it may be possible for more than one strategy to be the most effective for a given attacker profile.

\textit{At the end of this step, the analyst will obtain at least one strategy that best protects against the considered attacker profile and that corresponds to the suggested security control combinations to be implemented in the system.}  It is important to remember that this approach is a game. Therefore, as with any game, it is recommended that the game be re-constructed with different maximum budget values and re-played with different attacker profiles (as illustrated in Figure~\ref{fig:propMeth}). This can help gauge and compare the control combinations that should be used for the system under different constraints and goals.

% End Section

\section{CSAT: Control Selection Assistant Tool}
\label{sec:csat}
\begin{figure}[ht!]
    \centering
    \centerline{\includegraphics[width=\textwidth]{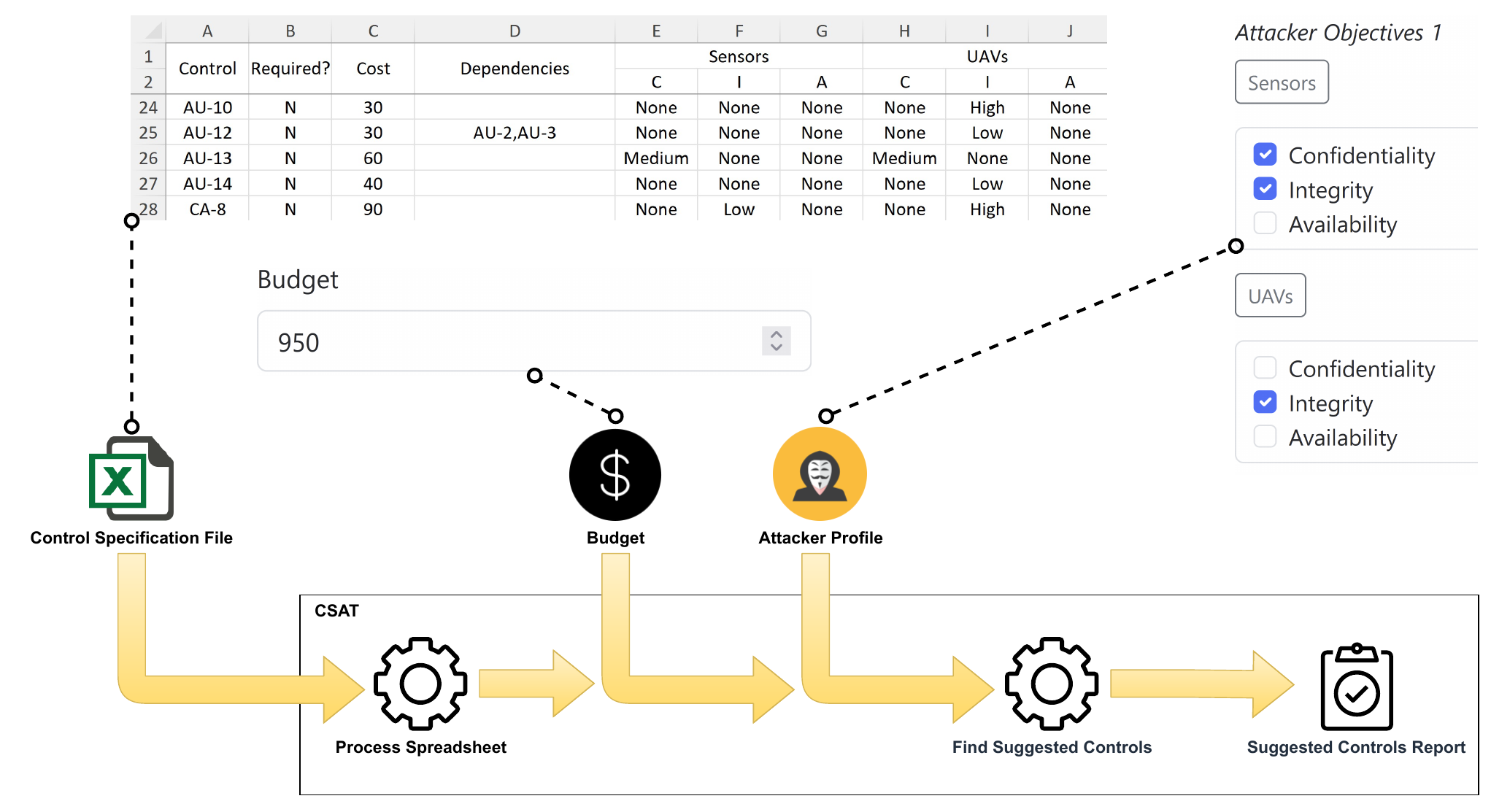}}
    \caption{Overview of CSAT's design and functionality}
    \label{fig:csatDesign}
\end{figure}

% Begin Section
In this section, we present the Control Selection Assistant Tool (CSAT)~\cite{FPS2024} 
% \footnote{CSAT is publicly available at the following link:  \url{https://gitlab.com/CyberSEA-Public/CSAT}.}
that automates the approach presented in Section~\ref{sec:approach}. The primary goal of CSAT is to eliminate the manual effort required to find  suggested control combinations for a given system. In fact, with~$N$ optional controls, there could exist up to~$2^N$  valid security control combinations for the approach, making it impractical when applied manually to large systems. With CSAT, results to the approach can be obtained without having found all valid security control combinations.
% Effectively, CSAT must solve a well-known NP complete problem known as the bounded 0-1 knapsack problem which can be described as follows: given a set of items with weight and value, what is the subset of unique items that will maximize value and be smaller than a given maximum weight~\cite{Vygen2001,Pan2018}. 
% As the first three steps of the approach are dependent on inputs from the security analyst, they cannot be automated by CSAT. However, the effort required to complete these steps is reasonable~\cite{GandALF2024}. 
% In fact, with $N$ optional controls, there could exist a total of $2^N$ combinations. With just 30 optional controls, ignoring dependencies and the cost constraints, this results in 1,073,741,824 combinations to generate. As a result, without such a tool the approach could not be used practically.
An overview of CSAT’s design and functionality can be seen in Figure~\ref{fig:csatDesign}.  
% CSAT performs two activities to find its suggested control combinations, which are denoted by gears in Figure~\ref{fig:csatDesign}. 
% Inputs and outputs to each of these activities are denoted with yellow arrows. 
The remainder of this section describes the components of this design. CSAT and the example data presented in this section are available at: \url{https://gitlab.com/CyberSEA-Public/CSAT}.

% \JJ{You should provide a URL to the tool and any additional specification files required for the examples, etc. This may not be the best spot for the link, but it is needed somewhere. }
% Available at: 

% \vspace{-1.5em}

\subsection{Control Specification File}

\begin{figure}[ht!]
\centering\centerline{\includegraphics[width=\textwidth]{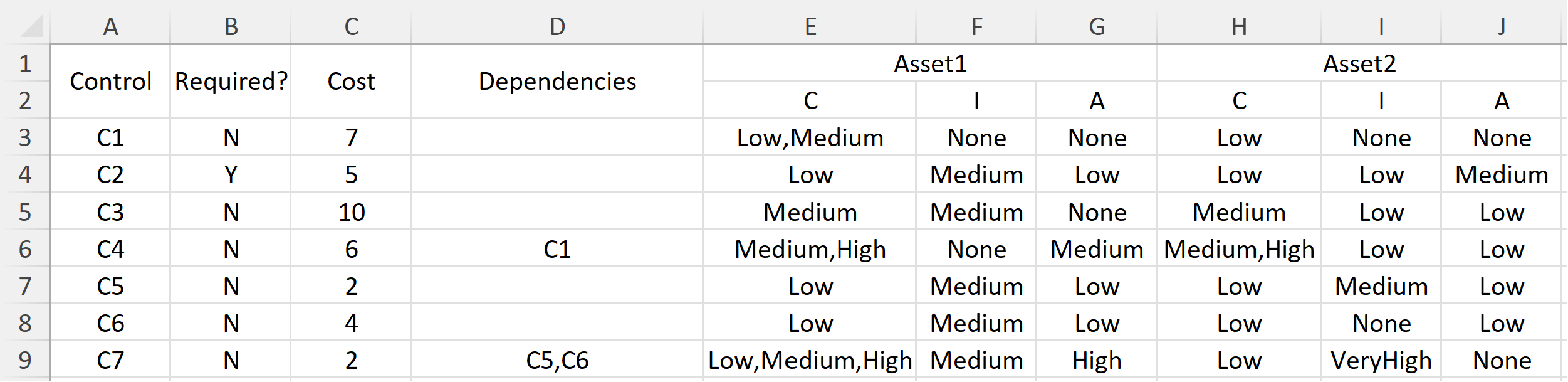}}
    \caption{An example of a filled CSAT control specification file}
    \label{fig:templateFilledCases}
\end{figure}

The atomic controls considered for the system, their effectiveness values, their cost, whether they are mandatory or optional,  and their dependencies, are provided through an Excel spreadsheet (\ie \texttt{.xlsx}) referred to as the control specification file.
% This information is known to the security analyst after completing \textbf{Steps} \textbf{1}, \textbf{2}, and \textbf{3}. 
An example of this spreadsheet being filled with controls can be seen in Figure~\ref{fig:templateFilledCases}.
% As shown in the figure, the control names are written in the first column. In the second column, a ``Y'' is used to denote a control as mandatory, while a ``N'' is used to denote a control that is optional. In the third column, the cost of each control is written. Note that units for each cost must be omitted.  Dependencies are written as a comma-separated list of controls in the fourth column. Lastly, the remaining columns represent the effectiveness each control has towards each security objective on each asset. 
% In the approach, the effectiveness of a control is defined as its ability to protect each security objective on each asset in the system.
As shown in the figure, CSAT currently only supports security objectives based around the CIA triad.  Additionally, each effectiveness value must be one of  \textit{None}, \textit{Low}, \textit{Medium}, \textit{High}, or \textit{VeryHigh}. These represent the normalized values of 0, 0.2, 0.5, 0.8, and 0.9 respectively. These values are adopted and adapted from the metrics in the Common Vulnerability Scoring System (CVSS)~\cite{CVSS}. No rating was assigned to the value of 1 as it is unrealistic to expect a single control to fully protect a security objective. To account for uncertainty, multiple effectiveness values can be provided for a control's effectiveness towards an asset's security objective (\eg in the figure, \textit{Low} versus \textit{Medium} for C1's effectiveness towards the confidentiality of Asset1). 
% Note that we omit the equation for calculating the effectiveness of a control combination from this work as it is not required to understand the functionality of CSAT. 

\subsection{Process Spreadsheet}

     % When running CSAT, it will give the display seen in Figure~\ref{fig:toolUI}. The tool
     As CSAT is a Python-based web tool,  it is accessed through a web browser. The control specification file can simply be passed into the file picker on the initial page. 
     % An empty control specification template file, and an example completed control specification file, can also be downloaded from this initial page. When the control specification file is inputted, it can be submitted by clicking the ``Submit Control File'' button. 
     CSAT will parse the specification file provided using \texttt{pandas}~\cite{pandas} and store the information internally.  
     
     % and store the control information from the control specification file. 
     % This model is simply a collection of data structures to hold the control names, their cost, effectiveness values, and their dependencies.
    
    % \begin{figure}[ht!]
    %     \centering
    %     \includegraphics[width=0.65\textwidth]{Figures/gameToolUI.PNG}
    %     \caption{The CSAT user interface when starting tool \JJ{This figure is not helpful; it can be removed}}
    %     \label{fig:toolUI}
    % \end{figure}

    \subsection{Budget and Attacker Profile}

\begin{figure}[t!]
        \centering
        \includegraphics[width=0.6\textwidth]{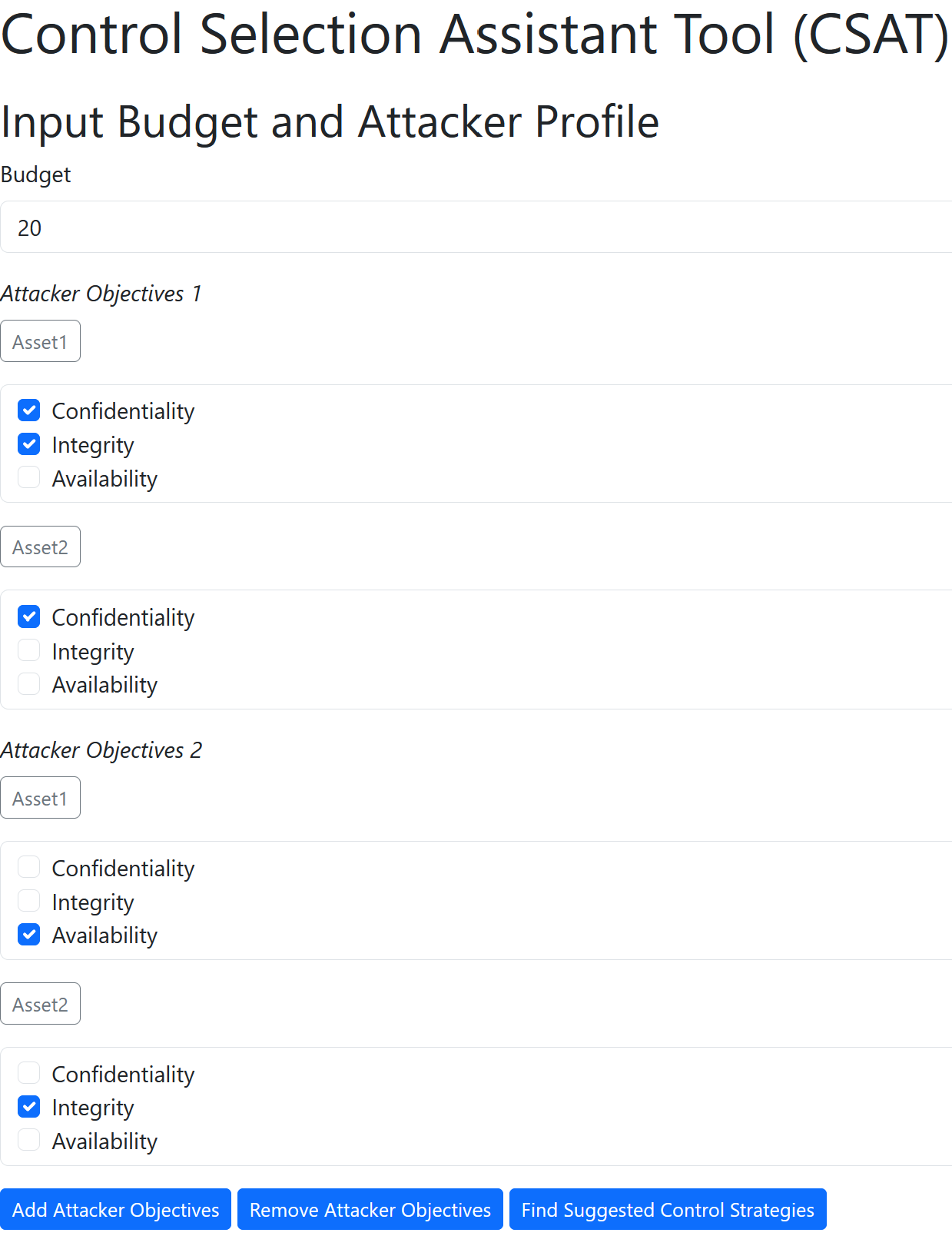}
        \caption{The CSAT user interface before the suggested controls are found}
        \label{fig:toolUIBeforeSub}
    \end{figure}

    After the control specification file is submitted and processed, a new page will be shown prompting for the budget and attacker profile. This can be seen in Figure~\ref{fig:toolUIBeforeSub}. 
    % In the approach, the budget is defined as the total cost that can be spent for the controls, while the attacker profile is defined as the expected attacker behaviour. Specifically, an attacker profile is established by considering and prioritizing different attacker objectives. Attacker objectives correspond to a set of security objectives for some assets that are equally expected to be targeted by an attacker. Within an attacker profile, several attacker objectives may be prioritized according to their perceived likelihood of being targeted by the attacker.  
    
    In the tool, the budget can simply be entered in the budget field. To specify the attacker profile, the security objectives in the first set of attacker objectives is specified by selecting the security objectives to be protected for each asset (each asset is a button on the tool which expands to its security objectives). Attacker objectives can be added and removed to create an attacker profile with as many ordered attacker objectives as desired. This is done using the ``Add Attacker Objectives'' button and the ``Remove Attacker Objectives'' button respectively. 

    % \begin{figure}[ht!]
    %     \centering
    %     \includegraphics[width=0.65\textwidth]{Figures/gameToolUiAfterSheet.png}
    %     \caption{The CSAT user interface after spreadsheet submission}
    %     \label{fig:toolUIafterSheet}
    % \end{figure}

    In this figure, the budget is 20, and we have an attacker profile with two ordered attacker objectives: the first targets the confidentiality and integrity of Asset1 and the confidentiality of Asset2 equally, and the second targets the availability of Asset1 and the integrity of Asset2 equally. Once submitted, CSAT will parse and store the budget and attacker objectives provided and will begin finding the suggested security controls for the system.
    % the budget is simply stored as a numerical value. On the other hand, the provided attacker profile is stored in the model as a list of sets, where each set contains the respective security objectives in the priority order given. 

    % In other words, with this attacker profile, we expect the confidentiality and integrity of Asset1 along with the confidentiality of Asset2 to be attacked equally. From the solutions found, then the controls which best protects the availability of Asset1 and the integrity of Asset2 should be selected as the suggested solution. 
    
    % A screenshot of the interface before finding the suggested controls can be seen in Figure~\ref{fig:toolUIBeforeSub}.

    \subsection{Find Suggested Controls}
    \label{gameTool:design:logic}

    \begin{figure}[t]
        \centering
        \includegraphics[width=\textwidth]{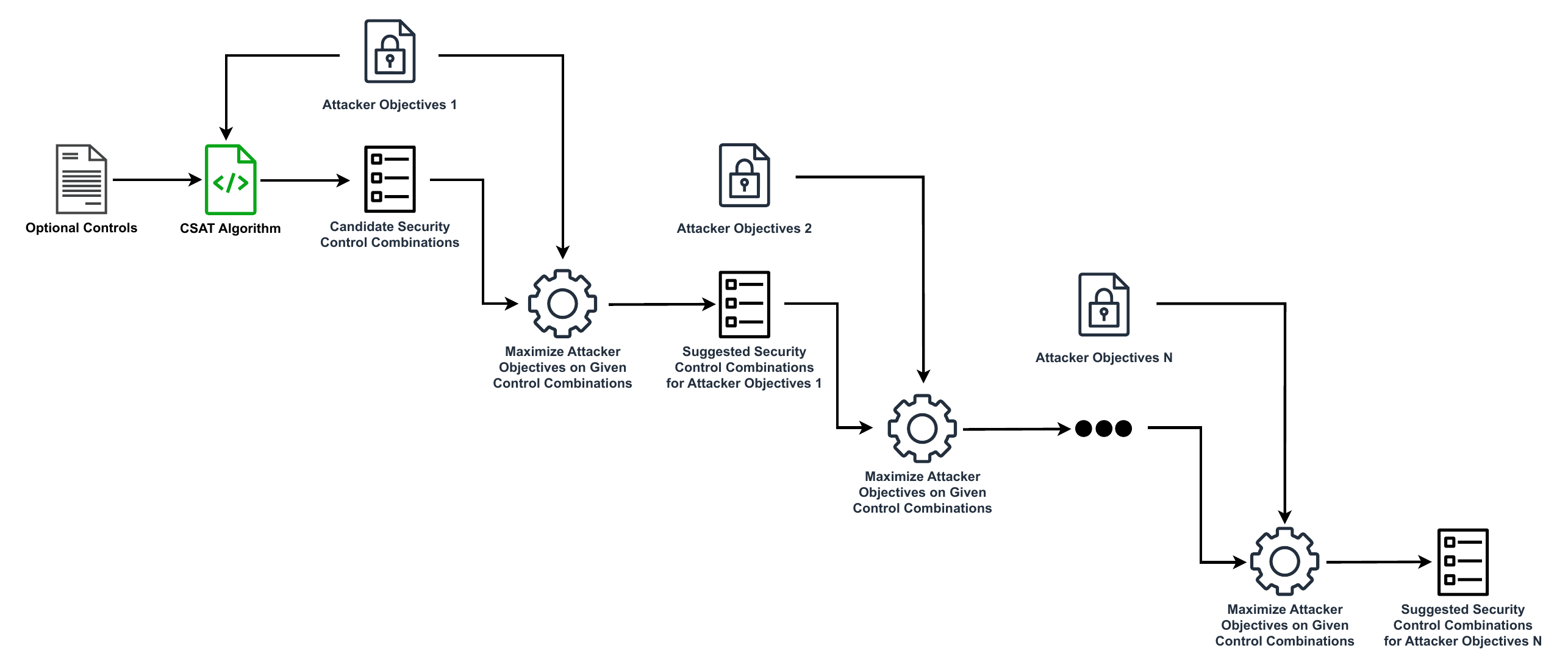}
        \caption{A visualization of the logic used by CSAT to find the suggested control combinations}
        \label{fig:algoViz}
    \end{figure}
   
    % To find the suggested control strategies, the ``Find Suggested Control Strategies'' button must be pressed. When the button is pressed, the provided budget and attacker profile are stored in the model. 
    
    % After CSAT processes and stores the budget and attacker profile, CSAT will calculate the suggested control combinations. 
    % In the approach, the suggested security control combinations are those maximize the effectiveness of each ordered objectives in the order  specified in the attacker profile.  
    The logic used by CSAT to find the suggested control combinations is visualized in Figure~\ref{fig:algoViz}. 
    % To find the suggested control combinations, 
    To begin, CSAT identifies a set of candidate security control combinations that are effective against the first set of attacker objectives. From this set of candidate security control combinations, CSAT will keep those which maximize the effectiveness towards the security objectives of the first set of attacker objectives. 
    % This will yield the most effective security control combinations for the first set of attacker objectives. 
    From these combinations, CSAT then keeps those which maximize the effectiveness towards the security objectives of the second set of attacker objectives. This process is repeated until there are no more ordered attacker objectives. 
    % Note that this logic is identical to that of the example provided in Figure~\ref{fig:combinedPersp}. A minor difference is that the logic used by CSAT first identifies candidate security control combinations. 
    Note that this logic is identical to that of the example provided in Figure~\ref{fig:combinedPersp}. A minor difference is that CSAT first identifies candidate security control combinations. Finding these candidate control combinations is essential for reducing the large number of security control combinations to evaluate, thus enabling the tool to generate its solutions within a reasonable timeframe.

    Finding these candidate control combinations corresponds to solving a well-known NP complete problem known as the bounded 0-1 knapsack problem~\cite{Vygen2001}. The algorithm developed to find these candidate control combinations (shown in in green in Figure~\ref{fig:algoViz}) has computational complexity $O(n)$ in special~cases and was previously described alongside the tool in~\cite{FPS2024}.

    % This logic may take time to run if a large number of security controls are specified in the control information spreadsheet. Because of this, a performance of the tool is provided

    \subsection{Suggested Security Control Combinations Report}

\begin{figure}[ht!]
        \centering\centerline{\includegraphics[width=\textwidth]{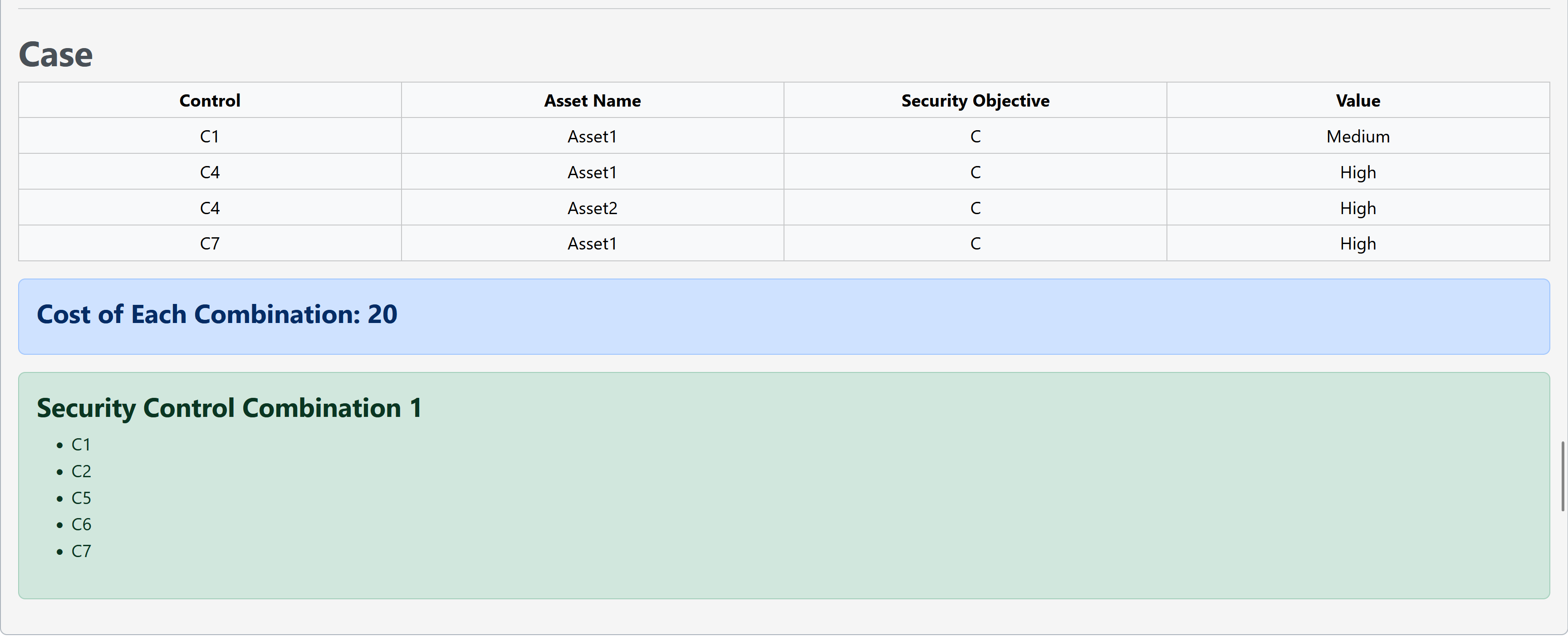}}
        \caption{Excerpt of a CSAT report}
        \label{fig:tollGuiAfterSubmitMultFirstRes}
    \end{figure}

    With the suggested control combinations found, CSAT will generate a report for the suggested control strategies. Figure~\ref{fig:tollGuiAfterSubmitMultFirstRes} shows part of the report generated by the tool after pressing submit for our running example.

    The ``Case'' heading is used to denote the suggested control combinations that were obtained for a particular case. As some controls were assigned multiple effectiveness values towards some of the security objectives, the suggested security controls must be found for each possible permutation of these uncertain values (which we define as a case). The case seen in the figure is the one were the confidentiality of C1 towards Asset~1 is \textit{Medium}, the confidentiality of C4 towards Asset~1 is \textit{High}, the confidentiality of C4 towards Asset~2 is \textit{High}, and the confidentiality of C7 towards Asset~1 is \textit{High}.

    In the report, the suggested security control combinations and their associated costs are displayed for each case. All suggested combinations for a case share the same cost as CSAT resolves ties in the results by selecting those with the lowest cost. As many cases could share the same suggested security controls, cases with identical results are printed sequentially. 
    % (omitted from the figure to save space). 
    % In this particular example, there were 21 cases for which the suggested security control combination is C1,~C2,~C5,~C6,~C7. However, three cases were found for which the suggested security control combination is C1,~C2,~C4,~C5.

% End Section

\section{Illustrative Example}
\label{sec:example}
% Begin Section

\begin{figure}[ht!]
        \centering
        \includegraphics[width=0.65\textwidth]{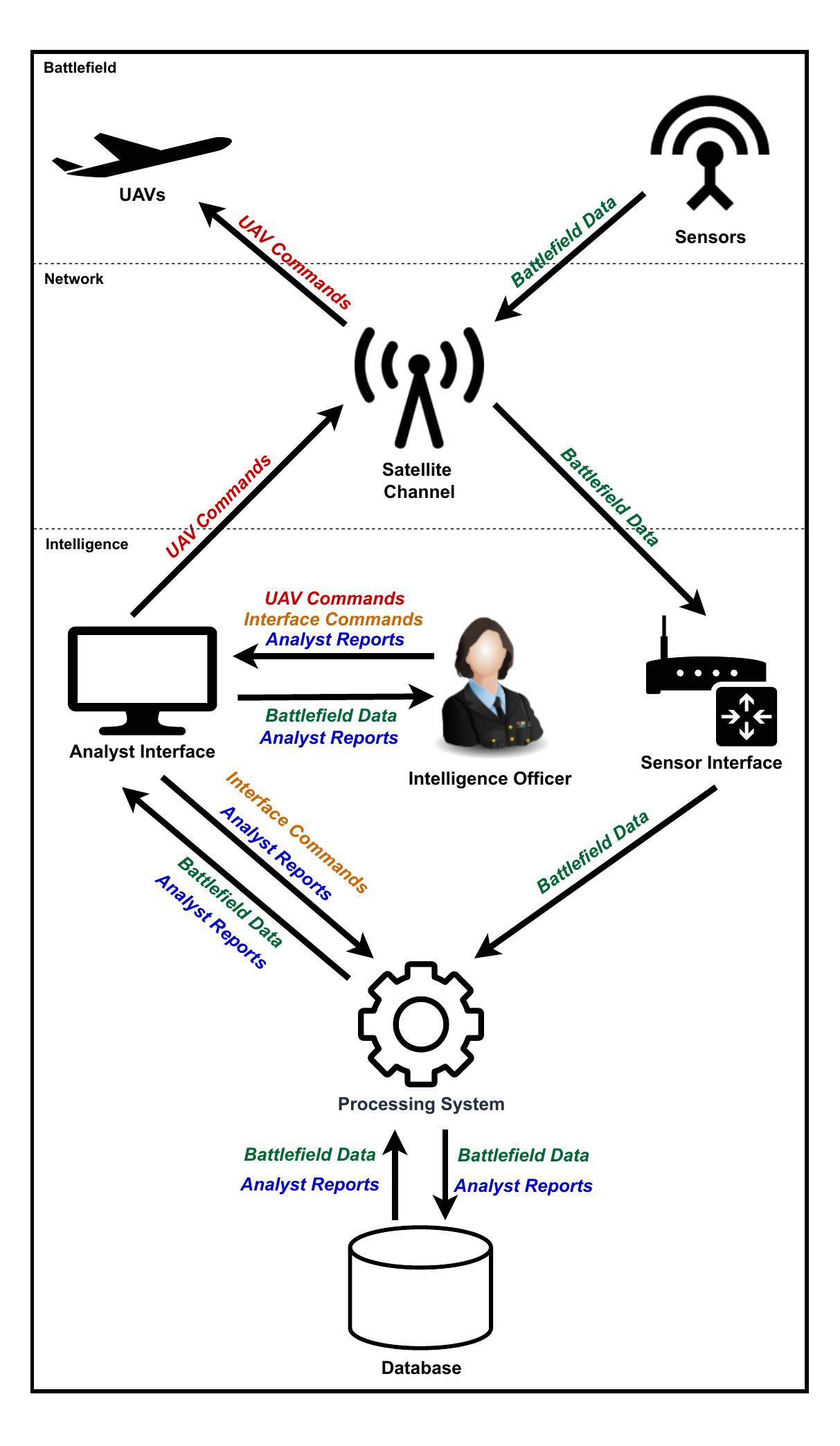}
        \caption{An overview of the \ravenclaw system architecture}
        \label{fig:basicArch}
    \end{figure}

In this section, we demonstrate how the approach presented in Section~\ref{sec:approach} could be applied to support the control selection activity for a large illustrative system. The system that will be used for this example is a fictional Internet of Things (IoT) system for the Canadian military called \ravenclaw. Its architecture has already been designed and can be seen in Figure~\ref{fig:basicArch}. Note that the arrows in this picture refer to the directions in which data flows in the system. 
% As this is a Canadian system, controls are selected from the ITSG-33 control catalogue. Additionally, the budget for \ravenclaw is \$950,000.

In this system, sensors collect battlefield information to be reported back for intelligence analysis. This data is processed with the processing system and stored in the database shortly after. An intelligence officer can then view the collected data from an analyst interface. Intelligence officers can also create and edit analysis reports from this interface, which are stored to the database when saved. Lastly, based on the analyses made, intelligence officers can send commands from this interface to UAVs in the battlefield.
    
The security analyst will apply the proposed approach to the \ravenclaw system as described in the following sections. 

\subsection{Identify Applicable Atomic Controls}

From the architecture provided in Figure~\ref{fig:basicArch}, we identify seven assets: the sensors, the UAVs, the satellite channel, the sensor interface, the analyst interface, the processing system, and the database. Note that although there are many sensors and UAVs, the sensors and UAVs will each be considered as a singular asset. This is because each sensor in the system is identical. The same can be said for the UAVs.

Because \ravenclaw is a Canadian government system, the analyst selects controls from the ITSG-33 control catalogue\footnote{ITSG-33 is the standard control catalogue to assist security practitioners in their efforts to protect information systems in compliance with applicable Government of Canada legislation, policies, directives, and standards~\cite{ITSG-33}.}. To comply with departmental requirements, it was decided by the organization that the following controls must be present in the system:
    \begin{itemize}
        \item \textit{AC-1: Access Control Policy and Procedures} 
        \item \textit{AC-3: Access Enforcement}
        \item \textit{AC-4: Information Flow Enforcement}
        \item \textit{AC-17: Remote Access}
        \item \textit{AC-23: Data Mining Protection}
        \item \textit{AU-1: Audit and Accountability Policy and Procedures}
        \item \textit{IA-3: Device Identification and Authentication }
        \item \textit{SC-8: Transmission Confidentiality and Integrity}
        
    \end{itemize}
\noindent %FIXED indent
As for the identified threats, the organization has provided to the security analyst the identified threats to each system asset using a  STRIDE threat model. The STRIDE model for the sensors can be seen in Table~\ref{tab:sensorsThreats}. To reduce cluttering this article, the STRIDE threat models for the other assets in the system can be viewed in the CSAT code repository linked above. Note that the organization has determined that the three security objectives of concern in the system are confidentiality, integrity, and availability. 

 \begin{table}[H] %FIXED Adjusted to booktabs
    \caption{STRIDE threat model for the \ravenclaw sensors}
        \label{tab:sensorsThreats}
        % \tiny
        % \tiny 
   \begin{center}

   % \scalebox{.75} {
    \resizebox{\textwidth}{!}{
    \begin{tabular}{lll}

    \multicolumn{1}{c}{\textbf{STRIDE Category}} & \multicolumn{1}{c}{\textbf{Threats}} & \multicolumn{1}{c}{\textbf{Security Objectives Violated}} \\ \toprule
    \multicolumn{1}{l}{Spoofing}               & \multicolumn{1}{p{3in}}{
    \parbox[t]{3in}{
    \begin{itemize}[itemsep=0pt, topsep=0pt, leftmargin=*]
        \item Data request from impersonating body
    \end{itemize}
    \strut}
    }             &     \parbox[t]{2.75in}{
    \begin{itemize}[itemsep=0pt, topsep=0pt, leftmargin=*]
        \item Confidentiality
    \end{itemize}
    \strut}          \\ \midrule
    \multicolumn{1}{l}{Tampering}               & \multicolumn{1}{p{3in}}{ \parbox[t]{3in}{
    \begin{itemize}[itemsep=0pt, topsep=0pt, leftmargin=*]
        \item Modification of device
        \item Outgoing data is modified
    \end{itemize}
    \strut}
    }             & 
    \parbox[t]{2.75in}{
    \begin{itemize}[itemsep=0pt, topsep=0pt, leftmargin=*]
        \item Integrity
    \end{itemize}
    \strut} 
    \\ \midrule
    \multicolumn{1}{l}{Repudiation}               & \multicolumn{1}{p{3in}}{N/A}             & N/A              \\ \midrule
    \multicolumn{1}{l}{Information Disclosure}               & \multicolumn{1}{p{3in}}{
    \parbox[t]{3in}{
    \begin{itemize}[itemsep=0pt, topsep=0pt, leftmargin=*]
        \item Outgoing data is read
    \end{itemize}
    \strut} 
    }             & 

    \parbox[t]{2.75in}{
    \begin{itemize}[itemsep=0pt, topsep=0pt, leftmargin=*]
        \item Confidentiality
    \end{itemize}
    \strut} 
    \\ \midrule
    \multicolumn{1}{l}{Denial of Service}               & \multicolumn{1}{p{3in}}{
    \parbox[t]{3in}{
    \begin{itemize}[itemsep=0pt, topsep=0pt, leftmargin=*]
        \item Physical destruction of device
    \end{itemize}
    \strut} 
    }             & 
    \parbox[t]{2.75in}{
    \begin{itemize}[itemsep=0pt, topsep=0pt, leftmargin=*]
        \item Availability
    \end{itemize}
    \strut} 
    \\ \midrule
    \multicolumn{1}{l}{Elevation of Privilege}               & \multicolumn{1}{p{3in}}{
    \parbox[t]{3in}{
    \begin{itemize}[itemsep=0pt, topsep=0pt, leftmargin=*]
        \item Improper SSH access
    \end{itemize}
    \strut} 
    }             &
    
    \parbox[t]{2.75in}{
    \begin{itemize}[itemsep=0pt, topsep=0pt, leftmargin=*]
        \item Confidentiality
        \item Integrity
    \end{itemize}
    \strut} 
    \\ \bottomrule
    \end{tabular}
    }
    
    \end{center}
        
    \end{table}

With the threats provided, the analyst can use the control catalogue to find relevant controls, and must ensure that each mandatory control is included at least once across all assets. All other applicable controls identified are therefore optional. The applicable controls for the sensors can be seen in Table~\ref{tab:sensorITSGControls}. The applicable controls for the other assets are found in the CSAT code repository. 

%FIXED to boooktabs and squished
{
\scriptsize
\begin{longtable}{p{1.1in} p{1.75in} p{2.725in}}
\caption{Applicable atomic controls for the \ravenclaw sensors}
\label{tab:sensorITSGControls} \\

\toprule
\multicolumn{1}{c}{\textbf{STRIDE Category}} & \multicolumn{1}{c}{\textbf{Threats}} & \multicolumn{1}{c}{\textbf{ITSG-33 Controls}} \\
\midrule
\endfirsthead

\caption{Applicable atomic controls for the \ravenclaw sensors (continued)}\\
\toprule
\multicolumn{1}{c}{\textbf{STRIDE Category}} & \multicolumn{1}{c}{\textbf{Threats}} & \multicolumn{1}{c}{\textbf{ITSG-33 Controls}} \\
\midrule
\endhead

Spoofing &
\begin{itemize}[itemsep=0pt, topsep=0pt, leftmargin=*]
    \item Data request from impersonating body
\end{itemize} &
\begin{itemize}[itemsep=0pt, topsep=0pt, leftmargin=*]
    \item \textit{AC-4: Information Flow Enforcement}
    \item \textit{AC-7: Unsuccessful Login Attempts}
    \item \textit{AC-17: Remote Access}
    \item \textit{AC-18: Wireless Access}
    \item \textit{IA-2: Identification and Authentication (Organizational Users)}
    \item \textit{SC-40: Wireless Link Protection}
\end{itemize} \\ \midrule

Tampering &
\begin{itemize}[itemsep=0pt, topsep=0pt, leftmargin=*]
    \item Modification of device
    \item Outgoing data is modified
\end{itemize} &
\begin{itemize}[itemsep=0pt, topsep=0pt, leftmargin=*]
    \item \textit{PE-1: Physical and Environmental Protection Policy and Procedures}
    \item \textit{PE-3: Physical Access Control}
    \item \textit{PE-6: Monitoring Physical Access}
    \item \textit{PE-20: Asset Monitoring and Tracking}
    \item \textit{SA-18: Tamper Resistance and Detection}
    \item \textit{SC-8: Transmission Confidentiality and Integrity}
    \item \textit{SI-7: Software, Firmware, and Information Integrity}
\end{itemize} \\ \midrule

Repudiation & N/A & N/A \\ \midrule

Information Disclosure &
\begin{itemize}[itemsep=0pt, topsep=0pt, leftmargin=*]
    \item Outgoing data is read
\end{itemize} &
\begin{itemize}[itemsep=0pt, topsep=0pt, leftmargin=*]
    \item \textit{AU-13: Monitoring for Information Disclosure}
    \item \textit{SC-8: Transmission Confidentiality and Integrity}
    \item \textit{SC-12: Cryptographic Key Establishment and Management}
\end{itemize} \\ \midrule

Denial of Service &
\begin{itemize}[itemsep=0pt, topsep=0pt, leftmargin=*]
    \item Physical destruction of device
\end{itemize} &
\begin{itemize}[itemsep=0pt, topsep=0pt, leftmargin=*]
    \item \textit{PE-1: Physical and Environmental Protection Policy and Procedures}
    \item \textit{PE-2: Physical Access Authorizations}
    \item \textit{PE-3: Physical Access Control}
    \item \textit{PE-6: Monitoring Physical Access}
    \item \textit{PE-20: Asset Monitoring and Tracking}
    \item \textit{SC-40: Wireless Link Protection}
\end{itemize} \\ \midrule

Elevation of Privilege &
\begin{itemize}[itemsep=0pt, topsep=0pt, leftmargin=*]
    \item Improper SSH access
\end{itemize} &
\begin{itemize}[itemsep=0pt, topsep=0pt, leftmargin=*]
    \item \textit{AC-6: Least Privilege}
    \item \textit{AC-7: Unsuccessful Login Attempts}
    \item \textit{AC-12: Session Termination}
    \item \textit{AC-24: Access Control Decisions}
    \item \textit{AU-2: Auditable Events}
    \item \textit{AU-8: Time Stamps}
    \item \textit{CA-8: Penetration Testing}
    \item \textit{IA-3: Device Identification and Authentication}
\end{itemize} \\ \bottomrule

\end{longtable}}

\subsection{Assign Effectiveness to Atomic Controls}

The analyst must now assign the effectiveness values for each identified applicable atomic control at mitigating the threats and protecting the security objectives listed in Table~\ref{tab:sensorsThreats}. The analyst has elected to assign qualitative ratings for the effectiveness of each atomic control that are mapped to quantitative values. These values, \textit{None} (0.0), \textit{Low} (0.2), \textit{Medium} (0.5), \textit{High} (0.8), and \textit{Very High} (0.9), align with the effectiveness values used for CSAT.

With the applicable controls gathered, the atomic payoff matrix can be created. Since there are many controls and assets, the atomic payoff matrix shall be broken into smaller matrices for readability; one for each asset. Using the metrics previously described, the analyst creates the atomic payoff matrix for the sensors. This matrix can be seen in Table~\ref{tab:atomicPayoffSensors} The matrices for the other assets are found in the CSAT code repository. 
 %FIXED to booktabs
{\scriptsize
    \begin{longtable}{llll}
    \caption{Atomic payoff matrix for the \ravenclaw sensors}
    \label{tab:atomicPayoffSensors} \\

                                & \multicolumn{3}{c}{\textit{Sensors}}                         \\ \cmidrule{2-4}
                                & \multicolumn{1}{c}{\textit{C}} & \multicolumn{1}{c}{\textit{I}} & \multicolumn{1}{c}{\textit{A}} \\ \toprule
    \endfirsthead
    \caption{Atomic payoff matrix for the \ravenclaw sensors (continued)} \\

                                & \multicolumn{3}{c}{\textit{Sensors}}                         \\ \cmidrule{2-4}
                                & \multicolumn{1}{c}{\textit{C}} & \multicolumn{1}{c}{\textit{I}} & \multicolumn{1}{c}{\textit{A}} \\
                                \midrule
    
    \endhead
    
    \multicolumn{1}{l}{\textit{AC-1: Access Control Policy and Procedures}}  & \multicolumn{1}{l}{None}  & \multicolumn{1}{l}{None}  & None  \\
    \multicolumn{1}{l}{\textit{AC-2: Account Management}}  & \multicolumn{1}{l}{None}  & \multicolumn{1}{l}{None}  &  None \\
    \multicolumn{1}{l}{\textit{AC-3: Access Enforcement}}  & \multicolumn{1}{l}{None}  & \multicolumn{1}{l}{None}  &  None \\
    \multicolumn{1}{l}{\textit{AC-4:  Information Flow Enforcement}}  & \multicolumn{1}{l}{Medium}  & \multicolumn{1}{l}{None}  & None  \\
    \multicolumn{1}{l}{\textit{AC-5: Separation of Duties}}  & \multicolumn{1}{l}{None}  & \multicolumn{1}{l}{None}  & None  \\
    \multicolumn{1}{l}{\textit{AC-6: Least Privilege} }  & \multicolumn{1}{l}{High}  & \multicolumn{1}{l}{Medium}  &  None \\
    \multicolumn{1}{l}{\textit{AC-7: Unsuccessful Login Attempts}}  & \multicolumn{1}{l}{None}  & \multicolumn{1}{l}{Medium}  &  Low \\
    \multicolumn{1}{l}{\textit{AC-8: System Use Notification}} & \multicolumn{1}{l}{None}  & \multicolumn{1}{l}{None}  & None  \\
    \multicolumn{1}{l}{\textit{AC-9: Previous Logon (Access) Notification}}  & \multicolumn{1}{l}{None}  & \multicolumn{1}{l}{None}  & None  \\
    \multicolumn{1}{l}{\textit{AC-12: Session Termination} } & \multicolumn{1}{l}{None}  & \multicolumn{1}{l}{Low}  & None  \\
    \multicolumn{1}{l}{\textit{AC-17: Remote Access} } & \multicolumn{1}{l}{None}  & \multicolumn{1}{l}{Low}  & None  \\
    \multicolumn{1}{l}{\textit{AC-18: Wireless Access}} & \multicolumn{1}{l}{None}  & \multicolumn{1}{l}{Medium}  & None  \\
    \multicolumn{1}{l}{\textit{AC-23: Data Mining Protection}}  & \multicolumn{1}{l}{None}  & \multicolumn{1}{l}{None}  & None  \\
    \multicolumn{1}{l}{\textit{AC-24: Access Control Decisions}} & \multicolumn{1}{l}{Low}  & \multicolumn{1}{l}{Low}  & None  \\
    \midrule
    \multicolumn{1}{l}{ \textit{AT-2: Security Awareness}}  & \multicolumn{1}{l}{None}  & \multicolumn{1}{l}{None}  & None  \\
    \midrule
    \multicolumn{1}{l}{\textit{AU-1: Audit and Accountability Policy and Procedures}} & \multicolumn{1}{l}{None}  & \multicolumn{1}{l}{None}  & None  \\
    \multicolumn{1}{l}{\textit{AU-2: Auditable Events}} & \multicolumn{1}{l}{None}  & \multicolumn{1}{l}{Low,Medium}  & None  \\
     \multicolumn{1}{l}{\textit{AU-3: Content of Audit Records}} & \multicolumn{1}{l}{None}  & \multicolumn{1}{l}{None}  & None  \\
     \multicolumn{1}{l}{\textit{AU-6: Audit Review, Analysis, and Reporting}} & \multicolumn{1}{l}{None}  & \multicolumn{1}{l}{None}  & None  \\
    \multicolumn{1}{l}{\textit{AU-8: Time Stamps}} & \multicolumn{1}{l}{None}  & \multicolumn{1}{l}{Low}  & None  \\
    \multicolumn{1}{l}{\textit{AU-9: Protection of Audit Information}} & \multicolumn{1}{l}{None}  & \multicolumn{1}{l}{None}  & None  \\
    \multicolumn{1}{l}{\textit{AU-10: Non-Repudiation}} & \multicolumn{1}{l}{None}  & \multicolumn{1}{l}{None}  & None  \\
    \multicolumn{1}{l}{\textit{AU-12: Audit Generation}} & \multicolumn{1}{l}{None}  & \multicolumn{1}{l}{None}  & None  \\
     \multicolumn{1}{l}{\textit{AU-13: Monitoring for Information Disclosure}} & \multicolumn{1}{l}{Medium}  & \multicolumn{1}{l}{None}  & None  \\
     \multicolumn{1}{l}{\textit{AU-14: Session Audit}} & \multicolumn{1}{l}{None}  & \multicolumn{1}{l}{None}  & None  \\
     \midrule
    \multicolumn{1}{l}{\textit{CA-8: Penetration Testing}} & \multicolumn{1}{l}{None}  & \multicolumn{1}{l}{Low}  & None  \\
    \midrule
    \multicolumn{1}{l}{\textit{IA-1: Identification and Authentication Policy and Procedures}} & \multicolumn{1}{l}{None}  & \multicolumn{1}{l}{None}  & None  \\
    \multicolumn{1}{l}{\textit{IA-2: Identification and Authentication (Organizational Users) }} & \multicolumn{1}{l}{None}  & \multicolumn{1}{l}{VeryHigh}  & None  \\
    \multicolumn{1}{l}{\textit{IA-3: Device Identification and Authentication} } & \multicolumn{1}{l}{None}  & \multicolumn{1}{l}{Medium}  & None  \\
    \multicolumn{1}{l}{\textit{IA-10: Adaptive Identification and Authentication}}  & \multicolumn{1}{l}{None}  & \multicolumn{1}{l}{None}  & None  \\
    \midrule
    \multicolumn{1}{l}{\textit{PE-1: Physical and Environmental Protection Policy and Procedures}} & \multicolumn{1}{l}{None}  & \multicolumn{1}{l}{None}  & None  \\
    \multicolumn{1}{l}{ \textit{PE-2: Physical Access Authorizations}} & \multicolumn{1}{l}{None}  & \multicolumn{1}{l}{Low}  & Low  \\
    \multicolumn{1}{l}{\textit{PE-3: Physical Access Control}} & \multicolumn{1}{l}{None}  & \multicolumn{1}{l}{Medium}  & High  \\
    \multicolumn{1}{l}{\textit{PE-6: Monitoring Physical Access} } & \multicolumn{1}{l}{None}  & \multicolumn{1}{l}{Medium}  & High  \\
    \multicolumn{1}{l}{\textit{PE-20: Asset Monitoring and Tracking} } &
    \multicolumn{1}{l}{None}  & \multicolumn{1}{l}{Low}  & Low  \\
    \midrule
    \multicolumn{1}{l}{\textit{SA-18: Tamper Resistance and Detection}} & \multicolumn{1}{l}{None}  & \multicolumn{1}{l}{High}  & None  \\
    \midrule
     \multicolumn{1}{l}{\textit{SC-5: Denial of Service Protection}} & \multicolumn{1}{l}{None}  & \multicolumn{1}{l}{None}  & None  \\
    \multicolumn{1}{l}{\textit{SC-7: Boundary Protection}} & \multicolumn{1}{l}{None}  & \multicolumn{1}{l}{None}  & None  \\
    \multicolumn{1}{l}{\textit{SC-8: Transmission Confidentiality and Integrity}} & \multicolumn{1}{l}{VeryHigh}  & \multicolumn{1}{l}{High}  & None  \\
    \multicolumn{1}{l}{\textit{SC-12: Cryptographic Key Establishment and Management}} & \multicolumn{1}{l}{High}  & \multicolumn{1}{l}{None}  & None  \\
    \multicolumn{1}{l}{\textit{SC-40: Wireless Link Protection} } & \multicolumn{1}{l}{Medium}  & \multicolumn{1}{l}{Medium}  & Medium  \\
    \midrule
    \multicolumn{1}{l}{ \textit{SI-4: Information System Monitoring}} & \multicolumn{1}{l}{None}  & \multicolumn{1}{l}{None}  & None  \\
    \multicolumn{1}{l}{\textit{SI-5: Security Alerts, Advisories, and Directives}}  & \multicolumn{1}{l}{None}  & \multicolumn{1}{l}{None}  &  None \\
    \multicolumn{1}{l}{\textit{SI-6: Security Functional Verification}} & \multicolumn{1}{l}{None}  & \multicolumn{1}{l}{None}  & None  \\
    \multicolumn{1}{l}{\textit{SI-7: Software, Firmware, and Information Integrity} } & \multicolumn{1}{l}{None}  & \multicolumn{1}{l}{High}  & None  \\
    \multicolumn{1}{l}{\textit{SI-10: Information Input Validation}}  & \multicolumn{1}{l}{None}  & \multicolumn{1}{l}{None}  & None  \\
    \multicolumn{1}{l}{\textit{SI-15: Information Input Filtering}}  & \multicolumn{1}{l}{None}  & \multicolumn{1}{l}{None}  & None  \\
    \bottomrule
    \end{longtable}}

\subsection{Assign Cost to Atomic Controls}
The cost of each atomic control can be seen in Table~\ref{tab:atomicControlCosts}. Note that the costs were assigned in consultation with the system designers. Also note that the costs below are for all the controls identified, not just those of the sensors.

    % Please add the following required packages to your document preamble:
% \usepackage[normalem]{ulem}
% \useunder{\uline}{\ul}{}
\begin{table}[t!]
    \caption{Atomic control costs for \ravenclaw}
    \label{tab:atomicControlCosts}
    \tiny

   \begin{center}
   \scalebox{1.2} {
\begin{tabular}{lc}

\multicolumn{1}{c}{\textbf{Control}}                                    & \textbf{Cost} \\
\toprule
\textit{AC-1: Access Control Policy and Procedures}                        & \$20,000         \\
\textit{AC-2: Account Management}                                          & \$60,000           \\
\textit{AC-3: Access Enforcement}                                          & \$40,000         \\
\textit{AC-4: Information Flow Enforcement}                                & \$30,000          \\
\textit{AC-5: Separation of Duties}                                        & \$10,000          \\
\textit{AC-6: Least Privilege}                                             & \$30,000          \\
\textit{AC-7: Unsuccessful Login Attempts}                                 & \$10,000          \\
\textit{AC-8: System Use Notification}                                     & \$20,000         \\
\textit{AC-9: Previous Logon (Access) Notification}                        & \$10,000          \\
\textit{AC-12: Session Termination}                                        & \$30,000           \\
\textit{AC-17: Remote Access}                                              & \$50,000          \\
\textit{AC-18: Wireless Access}                                            & \$30,000         \\
\textit{AC-23: Data Mining Protection}                                   & \$50,000         \\
\textit{AC-24: Access Control Decisions}                                   & \$20,000         \\
\midrule
\textit{AT-2: Security Awareness}                                          & \$200,000         \\
\midrule
\textit{AU-1: Audit and Accountability Policy and Procedures}              & \$20,000         \\
\textit{AU-2: Auditable Events}                                            & \$40,000         \\
\textit{AU-3: Content of Audit Records}                                    & \$30,000         \\
\textit{AU-6: Audit Review, Analysis, and Reporting}                       & \$20,000         \\
\textit{AU-8: Time Stamps}                                                 & \$10,000          \\
\textit{AU-9: Protection of Audit Information}                             & \$30,000         \\
\textit{AU-10: Non-Repudiation}                                            & \$30,000         \\
\textit{AU-12: Audit Generation}                                           & \$30,000         \\
\textit{AU-13: Monitoring for Information Disclosure}                      & \$60,000         \\
\textit{AU-14: Session Audit}                                              & \$40,000         \\
\midrule
\textit{CA-8: Penetration Testing}                                         & \$90,000         \\
\midrule
\textit{IA-1: Identification and Authentication Policy and Procedures}     & \$30,000        \\
\textit{IA-2: Identification and Authentication (Organizational Users)}    & \$30,000         \\
\textit{IA-3: Device Identification and Authentication}                    & \$30,000        \\
\textit{IA-10: Adaptive Identification and Authentication}                 & \$40,000          \\
\midrule
\textit{PE-1: Physical and Environmental Protection Policy and Procedures} & \$10,000       \\
\textit{PE-2: Physical Access Authorizations}                              & \$40,000         \\
\textit{PE-3: Physical Access Control}                                     & \$50,000          \\
\textit{PE-6: Monitoring Physical Access}                                  & \$80,000          \\
\textit{PE-20: Asset Monitoring and Tracking}                              & \$100,000        \\
\midrule
\textit{SA-18: Tamper Resistance and Detection}                            & \$100,000        \\
\midrule
\textit{SC-5: Denial of Service Protection}                                & \$40,000         \\
\textit{SC-7: Boundary Protection}                                         & \$60,000         \\
\textit{SC-8: Transmission Confidentiality and Integrity}                  & \$30,000          \\
\textit{SC-12: Cryptographic Key Establishment and Management}             & \$30,000         \\
\textit{SC-40: Wireless Link Protection}                                   & \$50,000          \\
\midrule
\textit{SI-4: Information System Monitoring}                               & \$60,000          \\
\textit{SI-5: Security Alerts, Advisories, and Directives}                 & \$20,000         \\
\textit{SI-6: Security Functional Verification}                            & \$70,000         \\
\textit{SI-7: Software, Firmware, and Information Integrity}               & \$40,000          \\
\textit{SI-10: Information Input Validation}                               & \$20,000         \\
\textit{SI-15: Information Input Filtering}                                & \$20,000        \\
\bottomrule
\end{tabular}}
\end{center}
\end{table}

For this example, each cost includes the effort required for an atomic control to be implemented in \ravenclaw. This is why policy controls have a cost associated to them (as they require an effort to develop). Note that the physical controls have a much higher cost than the other controls. This is because physical security will be more expensive as it requires physical measures to be purchased and installed (versus cyber controls which only require a software implementation). Also note that the \textit{AT-2: Security Awareness} control is expensive as well as much effort is required to develop a training course and provide this training to employees. Lastly, note that the organization has allocated a total budget of $B = \$ 950,000$.

\subsection{Specify and Generate Valid Control Combinations}
Next, the analyst must determine the valid security control combinations that could be considered for the system. To do this, they use security control algebra to specify the security control family from the mandatory and optional atomic controls identified in the previous steps. 
% Recall that the mandatory controls were provided at the start of Section~\ref{caseStudy:selectingControls}. The other applicable controls selected in Section~\ref{caseStudy:selectingControls} are therefore optional. 
Denoting the security control family as $F$, the security control family for this case study is specified as the following security control algebra term.
        
        \begin{eqnarray*}
        F &=& \textit{AC-1} \odot \textit{AC-3} \odot \textit{AC-4} \odot \textit{AC-17} \odot \textit{AC-23} \odot \textit{AU-1} \odot \textit{IA-3} \odot \textit{SC-8} \odot \\
        & & \opt{\textit{AC-2},\textit{AC-5},\textit{AC-6},\textit{AC-7},\textit{AC-8},\textit{AC-9},\textit{AC-12},\textit{AC-18},\\
        & &\textit{AC-24},\textit{AT-2},\textit{AU-2},\textit{AU-3},\textit{AU-6},\textit{AU-8},\textit{AU-9},\textit{AU-10},\\
        & & \textit{AU-12},\textit{AU-13},\textit{AU-14},\textit{CA-8},\textit{IA-1},\textit{IA-2},\textit{IA-10},\textit{PE-1}, \\
        & &
        \textit{PE-2},\textit{PE-3},\textit{PE-6},\textit{PE-20},\textit{SA-18},\textit{SC-5},\textit{SC-7},\textit{SC-12},\\
        & &\textit{SC-40},\textit{SI-4},\textit{SI-5},\textit{SI-6},\textit{SI-7},\textit{SI-10},\textit{SI-15}}
        \end{eqnarray*}

    The control names were omitted for clarity in the security control family above. For this case study, let's suppose the security analyst and organization have determined the dependencies shown in Table~\ref{tab:useCaseDeps}. This table presents the dependencies as requirement relations. It also presents the reasoning for which each dependency exists.  

    \begin{table}[t!]
    \caption{Atomic control dependencies for \ravenclaw}
    \label{tab:useCaseDeps}
\resizebox{0.95\textwidth}{!}{
    \begin{tabular}{lp{12cm}}

\multicolumn{1}{c}{\textbf{Dependency}}                                               & \multicolumn{1}{c}{\textbf{Reasoning for Dependency}}                                                                                    \\ \toprule
$\textit{AC-18} \xrightarrow[]{F} \textit{AC-17}$                                       & $\textit{AC-18}$ is a refinement of $\textit{AC-17}$                                                                                       \\ \midrule
$\textit{AU-2} \xrightarrow[]{F} \textit{AU-1}$                                         & $\textit{AU-2}$ requires an audit policy ($\textit{AU-1}$)                                                                                 \\ \midrule
$\textit{AU-3} \xrightarrow[]{F} \textit{AU-1}$                                         & $\textit{AU-3}$ requires an audit policy ($\textit{AU-1}$)                                                                                 \\ \midrule
$\textit{AU-6} \xrightarrow[]{F} \textit{AU-1}$                                         & $\textit{AU-6}$ requires an audit policy ($\textit{AU-1}$)                                                                                 \\ \midrule
$\textit{AU-8} \xrightarrow[]{F} \textit{AU-1}$                                         & $\textit{AU-8}$ requires an audit policy ($\textit{AU-1}$)                                                                                 \\ \midrule
$\textit{AU-9} \xrightarrow[]{F} \textit{AU-1}$                                         & $\textit{AU-9}$ requires an audit policy ($\textit{AU-1}$)                                                                                 \\ \midrule
$\textit{AU-10} \xrightarrow[]{F} \textit{AU-1}$                                        & $\textit{AU-10}$ requires an audit policy ($\textit{AU-1}$)                                                                                \\ \midrule
$\textit{AU-12} \xrightarrow[]{F} \textit{AU-1} \cdot \textit{AU-2} \cdot \textit{AU-3}$ & $\textit{AU-12}$ requires an audit policy ($\textit{AU-1}$), and depends on auditable events and content ($\textit{AU-2}$ and $\textit{AU-3}$) \\ \midrule
$\textit{AU-13} \xrightarrow[]{F} \textit{AU-1}$                                        & $\textit{AU-13}$ requires an audit policy ($\textit{AU-1}$)                                                                                \\ \midrule
$\textit{AU-14} \xrightarrow[]{F} \textit{AU-1}$                                        & $\textit{AU-14}$ requires an audit policy ($\textit{AU-1}$)                                                                                \\ \midrule
$\textit{IA-2} \xrightarrow[]{F} \textit{IA-1}$                                         & $\textit{IA-2}$ requires an authentication policy ($\textit{IA-1}$)                                                                        \\ \midrule
$\textit{IA-10} \xrightarrow[]{F} \textit{IA-2}$                                        & $\textit{IA-10}$ is an additional authentication mechanism to $\textit{IA-2}$                                                               \\ \midrule
$\textit{PE-2} \xrightarrow[]{F} \textit{PE-1}$                                         & $\textit{PE-2}$ requires a physical protection policy ($\textit{PE-1}$)                                                                    \\ \midrule
$\textit{PE-3} \xrightarrow[]{F} \textit{PE-1}$                                         & $\textit{PE-3}$ requires a physical protection policy ($\textit{PE-1}$)                                                                    \\ \midrule
$\textit{PE-6} \xrightarrow[]{F} \textit{PE-1}$                                         & $\textit{PE-6}$ requires a physical protection policy ($\textit{PE-1}$)                                                                    \\ \midrule
$\textit{PE-20} \xrightarrow[]{F} \textit{PE-1}$                                        & $\textit{PE-20}$ requires a physical protection policy ($\textit{PE-1}$)                                                                   \\ \midrule
$\textit{SC-12} \xrightarrow[]{F} \textit{SC-8}$                                        & Cryptographic key establishment/management ($\textit{SC-12}$) requires encryption to be employed ($\textit{SC-8}$)                         \\ \bottomrule
\end{tabular}
}

\end{table}

    At this point in the approach, we would typically expand the security control family $F$ to find all possible security control combinations, find those which respect the requirement relations, and filter the remaining combinations by the budget rule. This would be reasonable for small examples. However, it is not feasible to expand the  security control family above as the number of optional controls is very large; specifically, there are 39 optional controls. This security control family therefore generates $2^{39}$ possible security control combinations and cannot be found manually. Similarly, the requirement relations and budget rule cannot be applied manually as well. As a result, we will use CSAT to complete this illustrative example.

\subsection{Construct the Game Matrix}
Since the valid security control combinations could not be generated manually, and that these combinations represent the strategies of the security analyst in the game matrix, the game matrix cannot be constructed manually. However, to illustrate the general structure of this matrix, we have manually constructed a portion of this matrix in Table~\ref{tab:sampleMatrix} using two randomly selected control combinations. Only the sensor asset is included in the matrix as the inclusion of  all the assets would make it excessively wide. This again emphasizes the need to use CSAT to complete this example. 

\begin{table}[H]
    \caption{Sample atomic payoff matrix for \ravenclaw}
    \label{tab:sampleMatrix}
\resizebox{0.95\textwidth}{!}{
\begin{tabular}{llll}

                                                                                                                                                                                                                                                                                     & \multicolumn{3}{c}{\textit{Sensor}}                                         \\ \cmidrule{2-4}
                                                                                                                                                                                                                                                                                     & \multicolumn{1}{c}{\textit{C}}   & \multicolumn{1}{c}{\textit{I}}   & \multicolumn{1}{c}{\textit{A}} \\ \toprule
\multicolumn{1}{l}{\begin{tabular}[c]{@{}l@{}}$\textit{AC-1} \odot \textit{AC-2} \odot \textit{AC-3} \odot \textit{AC-4} \odot \textit{AC-5} \odot \textit{AC-6} \odot \textit{AC-7} \odot \textit{AC-12}  \mathrel{\odot} $ \\
$ \textit{AC-17} \odot \textit{AC-18} \odot \textit{AC-23} \odot \textit{AC-24}  \odot \textit{AU-1} \odot \textit{AU-2} \odot \textit{AU-3}  \mathrel{\odot}  $ \\ $     \textit{AU-6} \odot \textit{AU-8} \odot \textit{AU-10} \odot \textit{AU-12} \odot \textit{IA-1} \odot \textit{IA-2} \odot \textit{IA-3} \odot \textit{PE-1}  \mathrel{\odot} $ \\ $ \textit{PE-3} \odot \textit{SC-5} \odot \textit{SC-8} \odot \textit{SC-40} \odot \textit{SI-4} \odot \textit{SI-7}$ \end{tabular}} & \multicolumn{1}{c}{0.996}  & \multicolumn{1}{c}{0.99999} & 0.920                    \\ \midrule
\multicolumn{1}{l}{
\begin{tabular}[c]{@{}l@{}}$\textit{AC-1} \odot \textit{AC-2} \odot \textit{AC-3} \odot \textit{AC-4} \odot \textit{AC-5} \odot \textit{AC-6} \odot \textit{AC-7} \odot \mathrel{\textit{AC-8}} \odot $ \\
$ \textit{AC-12} \odot  \textit{AC-17} \odot \textit{AC-18} \odot 
        \textit{AC-23} \odot \textit{AC-24} \odot \textit{AU-1} \odot \mathrel{\textit{AU-2}} \odot  $ \\ $     \textit{AU-3} \odot \textit{AU-6} \odot \textit{AU-8} \odot \textit{AU-9} \odot \textit{AU-13} \odot \textit{AU-14} \odot \textit{IA-1} \odot \mathrel{\textit{IA-2}} \odot $ \\ $\textit{IA-3} \odot \textit{IA-10} \odot \textit{SC-8} \odot \textit{SI-4} \odot \textit{SI-5} \odot \textit{SI-10} \odot \textit{SI-15}$ \end{tabular}
 }                                                                                                                                                                                                                                                           & \multicolumn{1}{c}{0.996} & \multicolumn{1}{c}{0.99974} & \multicolumn{1}{c}{0.2}                    \\ \midrule
\multicolumn{1}{c}{...}                                                                                                                                                                                                                                                            & \multicolumn{1}{c}{...} & \multicolumn{1}{c}{...} & \multicolumn{1}{c}{...}                    \\ \bottomrule
\end{tabular}
}
\end{table}

\subsection{Play the Game}
We use CSAT to play the game. As mentioned in Section~\ref{sec:csat}, CSAT can find the suggested controls for a given attacker profile without having found all valid control combinations. The control specification file for this case study is publicly available in the CSAT project repository. Additionally, recall that the budget $B$ for \ravenclaw is  \$950,000.

To play the game, we will consider two different scenarios. Each scenario presents a different possible attacker profile for this system, and aims to show how the suggested controls for the system change based on this profile. These scenarios and their outcomes are described in the following subsections.

 \subsubsection{Scenario 1}

    This scenario considers an attacker profile where the attacker has three ordered attacker objectives to target: the confidentiality, integrity and availability of the sensor interface and analyst interface, followed by the confidentiality, integrity and availability of the processing system, and lastly followed by the confidentiality, integrity and availability of the database. 
    % Such attackers may spawn from the growth of advanced military-grade malware~\cite{malwareReport}.
    Using CSAT, we find that for all cases, there is but one suggested security control combination. This result has a cost of \$940,000. The suggested security control combination is as follows.

    {
    \small
 \begin{eqnarray*}
        \textit{Combo1.1} &=& \textit{AC-1} \odot \textit{AC-2} \odot \textit{AC-3} \odot \textit{AC-4} \odot \textit{AC-5} \odot \textit{AC-6} \odot \textit{AC-7} \odot \textit{AC-9} \odot \\
        & & \textit{AC-12} \odot  \textit{AC-17} \odot \textit{AC-18} \odot 
        \textit{AC-23} \odot \textit{AC-24} \odot \textit{AU-1} \odot \textit{AU-2} \odot  \\
        & & \textit{AU-3} \odot \textit{AU-8} \odot \textit{AU-10} \odot \textit{AU-12} \odot \textit{AU-14}  \odot \textit{IA-1} \odot \textit{IA-2} \odot  \textit{IA-3} \odot \\
        & & \textit{IA-10} \odot \textit{SC-5} \odot \textit{SC-8} \odot \textit{SC-12} \odot \textit{SI-4} \odot \textit{SI-7} \odot \textit{SI-10}
    \end{eqnarray*}
    }

    This combination contains a variety of  controls that protect the assets of concern. For example, \textit{IA-2: Identification and Authentication (Organizational Users)} protects a multitude of security objectives across multiple assets, namely the confidentiality and integrity of the sensor interface, analyst interface, and database. 
    Additionally, there are no physical controls in this combination which is logical as the expected attacker has no intention of targeting the physical assets of the system. 
    Note this combination considers 22 of the 39 applicable optional controls. Without CSAT, determining that this combination best protects against the expected attacker under the given constraints would have been infeasible. It took 10.659 seconds for CSAT to generate this result.
\subsubsection{Scenario 2}

    % There is concern that the data being sent from the sensors, and the data sent to the UAVs, be modified or jammed using advanced signal disruption technologies. From this description, 
    This scenario considers an attacker profile  where the attacker targets the integrity and availability of the sensors, UAVs, satellite channel, sensor interface, and analyst interface. 
    % Such an attacker is more prevalent with the rise in electronic warfare and jamming systems~\cite{electronicWarfare}.
    Using CSAT, we find there are different suggested security control combinations due to the analyst uncertainty. Specifically, we find one combination in four cases when \textit{AU-12: Audit Generation} has \textit{High} effectiveness towards the integrity of the analyst interface (\textit{Combo2.1}), and we find another in the other four cases when \textit{AU-12} has \textit{Medium} effectiveness towards the integrity of the analyst interface (\textit{Combo2.2}). Note that \textit{Combo2.1} has a cost of \$930,000, and \textit{Combo2.2} has a cost of \$940,000. The suggested security control combinations are as follows.
    
    {
    \small
    \begin{eqnarray*}
        \textit{Combo2.1} &=& \textit{AC-1} \odot \textit{AC-2} \odot \textit{AC-3} \odot \textit{AC-4} \odot \textit{AC-5} \odot \textit{AC-6} \odot \textit{AC-7} \odot \textit{AC-12} \odot \\
        & & \textit{AC-17} \odot \textit{AC-18} \odot \textit{AC-23} \odot \textit{AC-24}  \odot \textit{AU-1} \odot \textit{AU-2} \odot \textit{AU-3} \odot \\ 
        & & \textit{AU-6} \odot \textit{AU-8} \odot \textit{AU-10} \odot \textit{AU-12} \odot \textit{IA-1} \odot \textit{IA-2} \odot \textit{IA-3} \odot \textit{PE-1} \odot \\
        & & \textit{PE-3} \odot \textit{SC-5} \odot \textit{SC-8} \odot \textit{SC-40} \odot \textit{SI-4} \odot \textit{SI-7} \\
        \textit{Combo2.2} &=& \textit{AC-1} \odot \textit{AC-2} \odot \textit{AC-3} \odot \textit{AC-4} \odot \textit{AC-5} \odot \textit{AC-6} \odot \textit{AC-7} \odot \textit{AC-12}  \odot \\ 
        & & \textit{AC-17} \odot \textit{AC-18} \odot \textit{AC-23} \odot \textit{AC-24}  \odot \textit{AU-1} \odot \textit{AU-2} \odot \textit{AU-3} \odot \\
        & &  \textit{AU-6} \odot \textit{AU-8} \odot \textit{AU-10} \odot \textit{IA-1} \odot \textit{IA-2} \odot \textit{IA-3} \odot \textit{IA-10} \odot \textit{PE-1} \odot \\
        & & \textit{PE-3} \odot \textit{SC-5} \odot \textit{SC-8} \odot \textit{SC-40} \odot \textit{SI-4} \odot \textit{SI-7}
    \end{eqnarray*}
    
    }
    % \vspace{-1.5em}

    Both combinations differ by only one control; \textit{Combo2.1} has \textit{AU-12 Audit Generation} while \textit{Combo2.2}  has \textit{IA-10 Adaptive Identification and Authentication}. Although both controls are functionally different, and neither protect availability, they are both helpful in protecting the integrity of the targeted assets (audits and behavioural analyses can detect unauthorized data modifications).
    % In fact, if data is modified in transit by signal alteration technologies, audits and behavioural analyses can likely detect these changes.
    Hence, regardless of this uncertainty, we may say that both security control combinations are adequate in protecting against the expected attacker profile. 
    % As with \textbf{Scenario 1}, not all security objectives are protected fully. For example, \textit{AU-14 Session Audit} was disregarded despite having \textit{High} effectiveness towards the confidentiality of the database as the expected attacker is not interested in compromising this security objective. 
    In this scenario, the suggested combinations each consider 21 of the 39 applicable optional controls which, like \textit{Scenario 1}, would have been hard to identify without CSAT. Additionally, given that the uncertainty of an effectiveness value resulted in different suggested combinations for some of the cases, an analyst would have had to determine this manually by re-playing the game for each of the cases, which took CSAT just 16 minutes and 44 seconds to compute.

% This illustrative example and the corresponding scenarios highlight the fact that security control selection can indeed be seen as a game, and that the suggested controls to use depends greatly on the expected attacker profiles. 
% The proposed approach considers all the constraints and considerations required to perform control selection, and in contrast to existing approaches, also takes into consideration the expected attacker behaviours; an important factor to this problem that helps justify controls of interest and the risk to leave certain threats unaddressed. 

% End Section

\section{Discussion}
\label{sec:discussion}
% Begin Section

 Approaching security control selection as a game emphasizes the human element in deciding how to most effectively protect a system's assets under various considerations such as budgetary constraints. Selecting controls for a system is indeed a human-centric problem as the large number of potential controls to use from a control catalogue can be overwhelming and could lead to many mistakes in the chosen combination of security controls selected for the system. To expand on this point, selecting security controls exclusively on technical considerations while overlooking attacker behaviours is a fundamentally flawed approach in addressing this issue, as ultimately, it is humans which are conducting attacks.  Given that the proposed approach emphasizes the need to reflect on potential attacker behaviours, it prioritizes the human-centric aspect to find its solutions. Additionally, viewing this problem as a game captures the opposing dynamics of the attacker and analyst, aligning with the real-world motivations of both actors.

 Unfortunately, a limitation with many existing game-theoretic approaches from addressing cybersecurity challenges is their heavy reliance on assumptions. Specifically, game theory depends on assumptions on the players, such as the players knowing every strategy available to them, knowing the probabilities of every move, and knowing the payoff functions~\cite{OWEN2015573}. Compared to existing works, the proposed approach can be used practically as it does not rely on assigning probabilities for the likelihood of attacks succeeding. In fact, such probabilities would generally be unattainable in practice without sufficient historical data. Instead, the approach focuses on general assumptions about which security objectives could be violated by the attacker. Security analysts cannot realistically predict exact probabilities of attack, but they can make informed assumptions regarding which system components might be more attractive to attackers. These considerations ensure that the proposed approach is systematic, repeatable, and realistic, thereby minimizing the influence of human bias on the results of the game and eliminating many of the required assumptions with existing game-theoretic approaches. 

% While the proposed approach may seem limited by the security analyst's certainty in the effectiveness values for each atomic control, a sensitivity analysis was performed on the effectiveness metrics used in Section~\ref{sec:example} and revealed that the results of the approach were not sensitive to these values. The full details of this analysis are omitted due to space limitations. In any case, to allow the analyst to express their uncertainty, it is possible to allow them to provide multiple values when assigning the effectiveness for atomic controls as performed in CSAT.   
% Lastly, as atomic controls are considered indivisible components, we assume costs can be independently assigned to each atomic control. From a business perspective, this assumption may not always hold as the aggregation of certain controls could result in lower total costs to implement some control combinations. We argue that this limitation is unlikely to occur unless the controls are provided by third party vendors. In such cases, the cost function outlined in Definition~\ref{def:costInduction} can be adapted to combine costs in a different manner.

Although CSAT can automate the approach, finding the suggested security controls remains an NP-complete problem. Therefore, CSAT cannot be expected to generate its solutions rapidly. However, a performance analysis was conducted for a typical\footnote{A typical usage of CSAT refers to a scenario in which the optional controls considered include a balanced mix of controls with varying effectiveness and cost.} usage of CSAT and revealed that the tool has an upper limit capacity of 50 optional controls. With this number of optional controls, results were generated in approximately one hour. This running time is acceptable given that 50 optional controls is a significantly large number of inputs and that the results of CSAT are not needed in real-time. To support this claim, considering 50 controls from ITSG-33 is equivalent to considering 17\% of the controls from this catalogue (given that ITSG-33 has approximately 300 security controls~\cite{ITSG-33}).  To further support this claim, despite having identified 39 optional controls in Section~\ref{sec:example}, the results for both scenarios were generated in a reasonable amount of time. The approach can therefore be used practically towards large systems using CSAT.

\section{Conclusions and Future Work}
\label{sec:conclusion}
% Begin Section
Ensuring effective security controls are selected for a system can greatly impact its security. In this work, a game-theoretic approach to security control selection is proposed in which a game is played by a security analyst to determine security controls which best mitigate expected attacker profiles. The suggested controls can  help make a security analyst feel more confident in their decision to implement some controls over others. To demonstrate the approach's scalability for large systems, the approach was applied to a large fictional Canadian military system using a previously proposed tool~\cite{FPS2024}. Despite 39 optional controls being considered in this example, results were generated in reasonable time, demonstrating the scalability of the approach when automated and applied towards large systems.

In future work, we aim to investigate the outcomes of the game when played from the attacker perspective. As the game matrix used in this approach is known only to the security analyst, it is not realistic to play the game as an attacker. However, if the attacker were to gain access to the game matrix and wished to play the game, they would  make assumptions about possible security analyst behaviours and create \textit{defender profiles}, representing security control combinations that the security analyst is likely to choose for the system. The attacker could then play the game from the opposite perspective of the analyst, by selecting the security objectives that are least protected against security control combinations that might be chosen. While playing the game from this perspective may not be realistic, it could potentially identify weaknesses in certain security controls or pinpoint security objectives that are more likely to be successfully violated by an attacker (\ie weaker links in the system). Additionally, for future work, changes to CSAT, such as allowing for other formats than \texttt{.xlsx} for inputting control information, or improving its general look-and-feel, 
% (\eg colours, buttons, etc.)
 could be made to improve the usability of the tool. 

% In fact, a prototype of the tool has already been developed and is available at the following link: \url{https://github.com/DylanLeveille/CSAT}.
% End Section

\section*{Acknowledgment}
This research is funded in part by the Human-Centric Cybersecurity Partnership under the SSHRC Partnership Grants program.

  %% the following bibliography is gererated manually for the sake of brevity
  %% only; please use a separate .bib file in your submission

% \begin{thebibliography}{Kos97}

% % \bibitem[Kos97]{koslowski:mib}
% % J{\"u}rgen Koslowski.
% % \newblock Monads and interpolads in bicategories.
% % \newblock {\em Theory Appl. Categ.}, 3(8):182--212, 1997.

% \end{thebibliography}

% Bibliography
\bibliographystyle{alphaurl}
\bibliography{LMCS2024}

\end{document}